\documentclass{emulateapj}

\usepackage{multirow}
\usepackage{amsmath, amsthm, amssymb}
\usepackage{natbib}

\newcommand{\ud}{\mathrm{d}}

\shorttitle{Radiation Transfer of Models of Massive Star Formation. I.}
\shortauthors{Zhang \& Tan}

\begin{document}

\title{Radiation Transfer of Models of Massive Star Formation. I. Dependence on Basic Core Properties}

\author{Yichen Zhang}
\affil{Department of Astronomy, University of Florida, Gainesville, FL 32611, USA;\\ yczhang@astro.ufl.edu}
\author{Jonathan C. Tan}
\affil{Departments of Astronomy \& Physics, University of Florida, Gainesville, FL 32611, USA;\\ jt@astro.ufl.edu}

\begin{abstract}

Radiative transfer calculations of massive star formation are
presented. These are based on the Turbulent Core Model of McKee \& Tan
and self-consistently included a hydrostatic core, an inside-out
expansion wave, a zone of free-falling rotating collapse, wide-angle
dust-free outflow cavities, an active accretion disk, and a massive
protostar. For the first time for such models, an optically thick
inner gas disk extends inside the dust destruction front. This is
important to conserve the accretion energy naturally and for its
shielding effect on the outer region of the disk and envelope. The
simulation of radiation transfer is performed with the
Monte Carlo code of Whitney, yielding spectral energy distributions
(SEDs) for the model series, from the simplest spherical model to the
fiducial one, with the above components each added
step-by-step. Images are also presented in different wavebands of
various telescope cameras, including Spitzer IRAC and MIPS, SOFIA
FORCAST and Herschel PACS and SPIRE.  The existence of the optically
thick inner disk produces higher optical wavelength fluxes but reduces
near- and mid-IR emission. The presence of outflow cavities, the
inclination angle to the line of sight, and the thickness of the disk
all affect the SEDs and images significantly. For the high mass
surface density cores considered here, the mid-IR emission can be
dominated by the outflow cavity walls, as has been suggested by De
Buizer. The effect of varying the pressure of the
environment bounding the surface of the massive core is also studied. 
With lower
surface pressures, the core is larger, has lower extinction and
accretion rates, and the observed mid-IR flux from the disk can then
be relatively high even though the accretion luminosity is lower. In
this case the silicate absorption feature becomes prominent, in
contrast to higher density cores forming under higher pressures.

\end{abstract}


\section{Introduction}

``How do massive stars form?'' is still a debated question (e.g.,
\citealt[]{Beuther07}). One basic problem is massive protostars become
so luminous that radiation pressure may stop the accretion and growth
of the star. One possible mechanism to form a massive star can be
considered as a scaled-up version of low-mass star formation. If the
gas is turbulent and threaded by sufficiently strong magnetic fields
then fragmentation may be suppressed for cores much more massive than
the Jeans mass. The conditions leading to this suppression should be
relatively rare since massive stars are rare and make up only a small
mass fraction of the final star cluster. These massive cores are
expected to form from highly pressurized clumps of gas, in which case
they start with high densities, short free-fall times and therefore
high accretion rates. This is the basic scenario of the Turbulent Core
Model (\citealt[]{MT03}, hereafter MT03). 
Other radically different possibilities are that massive stars form
through stellar mergers (\citealt[]{Bonnell98}) or accrete most of
their mass from initially unbound material (competitive accretion
model, \citealt[]{Bonnell01}, \citealt[]{Bonnell04},
\citealt{Bate09a,Bate09b}).

Answering this question is difficult observationally because massive
star formation occurs in distant and highly obscured regions. However,
with the improving sensitivity and spatial resolution provided by the
instruments such as the {\it Herschel Space Telescope}, {\it
  Stratospheric Observatory for Infrared Astronomy (SOFIA)}, {\it Gran
  Telescopio Canarias (GTC) - CanariCam}, the {\it James Webb Space
  Telescope} and Thirty-Meter class telescopes, we expect to see a
faster advance in the research of massive star formation and hope this
question can be finally solved.

In order to interpret observations such as spectral energy
distributions (SEDs) or images, and then have a better understanding
of the properties of a massive protostar and its evolution, a number
of models have already been developed to fit or compare with
observations. For example, \citet[]{Robitaille06} have developed a
very impressive model grid containing 200,000 SEDs covering a large
parameter space, which is publicly available and now widely used
(referred to below as the ``Robitaille model'').  However, this grid
mainly covers low-mass young stellar objects (YSOs) and when massive
protostars are considered, they do not have the properties expected
for fiducial parameters of the turbulent core model. Also these models
do not consider the presence of an optically thick gaseous disk inside
the dust destruction front. \citet[]{Molinari08} used the same methods
employed by the Robitaille model but now for protostellar parameters
based on the turbulent core model, to developed a SED model grid for
massive YSOs. They found that the SED can be a diagnostic tool to
determine the evolutionary stage of a massive YSO. However, in this
work the representation of the turbulent core model is quite
approximate and limitations of the Robitaille model framework are
still present. \citet[]{CM05} developed an analytic solution for
far-IR SED of a protostar embedded in a spherically symmetric
molecular cloud, but this method could not allow for presence of
accretion disks and protostellar outflow cavities.
\citet[]{Indebetouw06} investigated the effects of clumpy structures
in the molecular envelope around massive protostars. However, again,
their models were not tuned to the parameters of the turbulent core
model: for example, they considered much more massive structures
(e.g. $\sim 5\times 10^4\;M_\odot$ contained in a sphere of radius
2.5 pc), more representative of star forming clumps that form entire
star clusters. They did not consider protostellar disks.

Our aim is thus to develop a new model of massive protostars, based on
the Turbulent Core Model of \citet{MT02} and MT03, including all
the important components self-consistently, and then perform
simulations of the radiation transfer to see whether different
components or evolutionary stages are represented in the SEDs and
images. Starting with a fiducial hydrostatic core of $60\;M_\odot$
bounded by the pressure of a self-gravitating clump of mean mass
surface density $\Sigma_{\rm cl}=1\:{\rm g\:cm^{-2}}$, we then
consider its appearance once an $8\;M_\odot$ protostar has formed at its
center. We develop a series of protostellar models of increasing
realism: starting with a simple hydrostatic core, we then apply the
inside-out expansion wave solution (\citealt[]{Shu77}), but
generalized to singular polytropic spheres (\citealt[]{MP97}) and the
free-fall rotating collapse solution by \citet[]{Ulrich}. A
circumstellar disk is expected to form around the protostar and this
is important to transfer angular momentum and to solve the radiation
pressure problem (e.g. \citealt[]{jijina96}, \citealt[]{Krumholz07}). 
Presently, there is evidence for rotating toroids around massive
protostars (e.g. \citealt[]{Beltran05}) but little direct evidence for
Keplerian protostellar disks, which will probably require the angular
resolution of ALMA.  We include the disk with an $\alpha$-disk model
(\citealt[]{SS73}). Unlike \citet[]{Robitaille06} and
\citet[]{Molinari08}, we include an optically thick inner disk with
gas opacities inside the dust destruction radius, which is important
to conserve the accretion energy naturally. Accretion is expected to
drive strong bipolar outflows and sweep up the material in the core
and form cavities. These outflow cavities have been observed around
massive protostars and may determine the mid-IR morphology (e.g.,
\citealt[]{deBuizer06}).

The assumptions of each component of our model and the model series
are introduced in detail in the next section. In
Section. \ref{sec:simulation}, we discuss our simulations, including
the Monte Carlo radiation transfer code, and the dust and gas
opacities we use. In Section. \ref{sec:results}, we present the SED
and image results of our models. In Section. \ref{sec:conclusions}, we
summarize our main results, including a comparison of our model with
other works. In future papers we will examine additional refinements,
especially the development and material content of the outflow
cavities, and present results of the evolutionary sequence of massive
star formation based on the fiducial model we have started to develop
here.

\section{Massive Protostar Model}

\subsection{Envelope}

\subsubsection{Hydrostatic Outer Envelope}
\label{sec:core}

Following MT03 and \citet[]{Tan08}, we define a
``star-forming core'' as a region of a molecular cloud that will form
a single star or close binary, and assume it is self-similar,
self-gravitating in near virial equilibrium and spherical. The density
and pressure each have power-law dependencies on radius, $\rho\propto
r^{-k_\rho}$ and $P\propto r^{-k_p}$. This smooth power-law density
distribution is only an approximation, especially given that massive
cores in virial equilibrium but with gas temperatures $\sim 10\;{\rm
  K}$ must be supported by non-thermal forms of pressure support, such
as turbulence and/or magnetic fields.  Clumpy substructures are likely
to form inside a turbulent core and this will affect radiative
transfer through the core - basically making it somewhat easier for
shorter wavelength photons to propagate through the
core. \citet[]{Indebetouw06} performed radiative transfer models of
clumpy cores. However, given the uncertain nature of the clumping, we
defer such considerations to a future paper, and first calculate the
properties of smoothly distributed gas and dust.

From the above power law distributions, it follows that the core is
polytropic with $P\propto \rho^{\gamma_p}$. The case with $\gamma_p=1$
is a singular isothermal sphere (\citealt[]{Shu77}) and in the other
limit $\gamma_p=0$ corresponds to a logotropic sphere
(\citealt{MP96,MP97}). For the models presented here,
we follow MT03 and adopt $k_\rho=1.5$, thus $\gamma_p=\frac{2}{3}$ and
$k_p=1$. The equation of hydrostatic equilibrium gives
\begin{equation}
M(r)=\frac{k_pc^2r}{G},
\end{equation}
and
\begin{equation}
\rho(r)=\frac{(3-k_\rho)(k_\rho-1)c^2}{2\pi G r^2}, \label{den}
\end{equation}
where $c=(P/\rho)^{1/2}$ is the effective sound speed.

We assume the total mass in the core is 60 $M_\odot$. If the efficiency
$\epsilon_{*f}$ is 0.5, which is estimated from low-mass cases
(\citealt[]{MM00}), a star with mass 
$m_{*f}=\epsilon_{*f}
M_\mathrm{core}=30\;M_\odot$ can finally form out of the core. For a
core with such mass, the core radius is (eq. (20) in MT03)
\begin{equation}
R_\mathrm{core}=0.057\Big(\frac{M_\mathrm{core}}{60\;M_\odot}\Big)^{1/2}\Sigma_\mathrm{cl}^{-1/2}\;\mathrm{pc},
\end{equation}
where $\Sigma_\mathrm{cl}$ is the mean surface density in the
molecular clump in which the core is embedded, and $1\;
\mathrm{g/cm}^{-2}$ is used as a fiducial value. We also consider
models with higher and lower values of $\Sigma_{\rm cl}$.

\subsubsection{Expansion Wave}
\label{sec:expansionwave}

\begin{figure*}
\begin{center}
\includegraphics[width=\columnwidth]{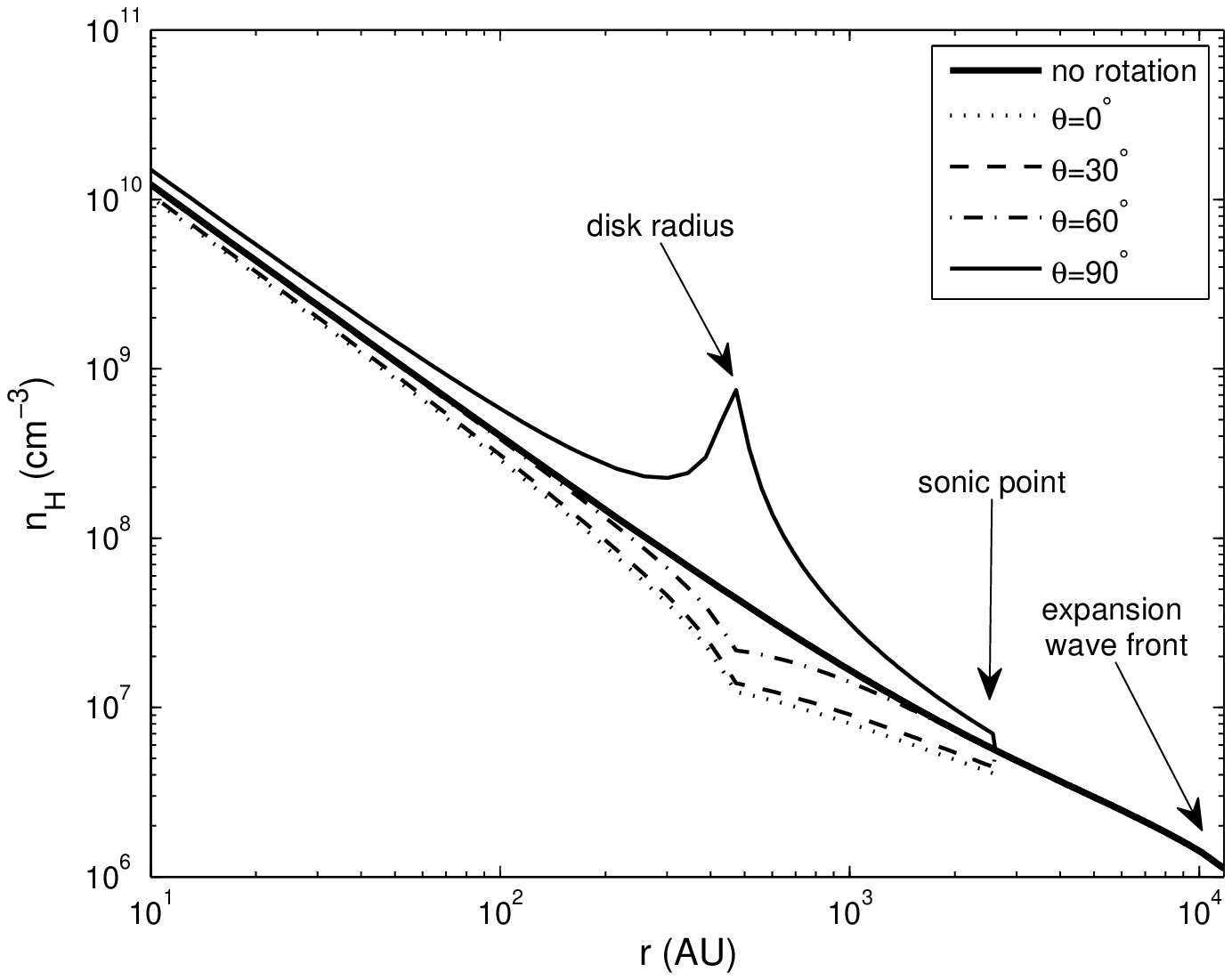}
\includegraphics[width=\columnwidth]{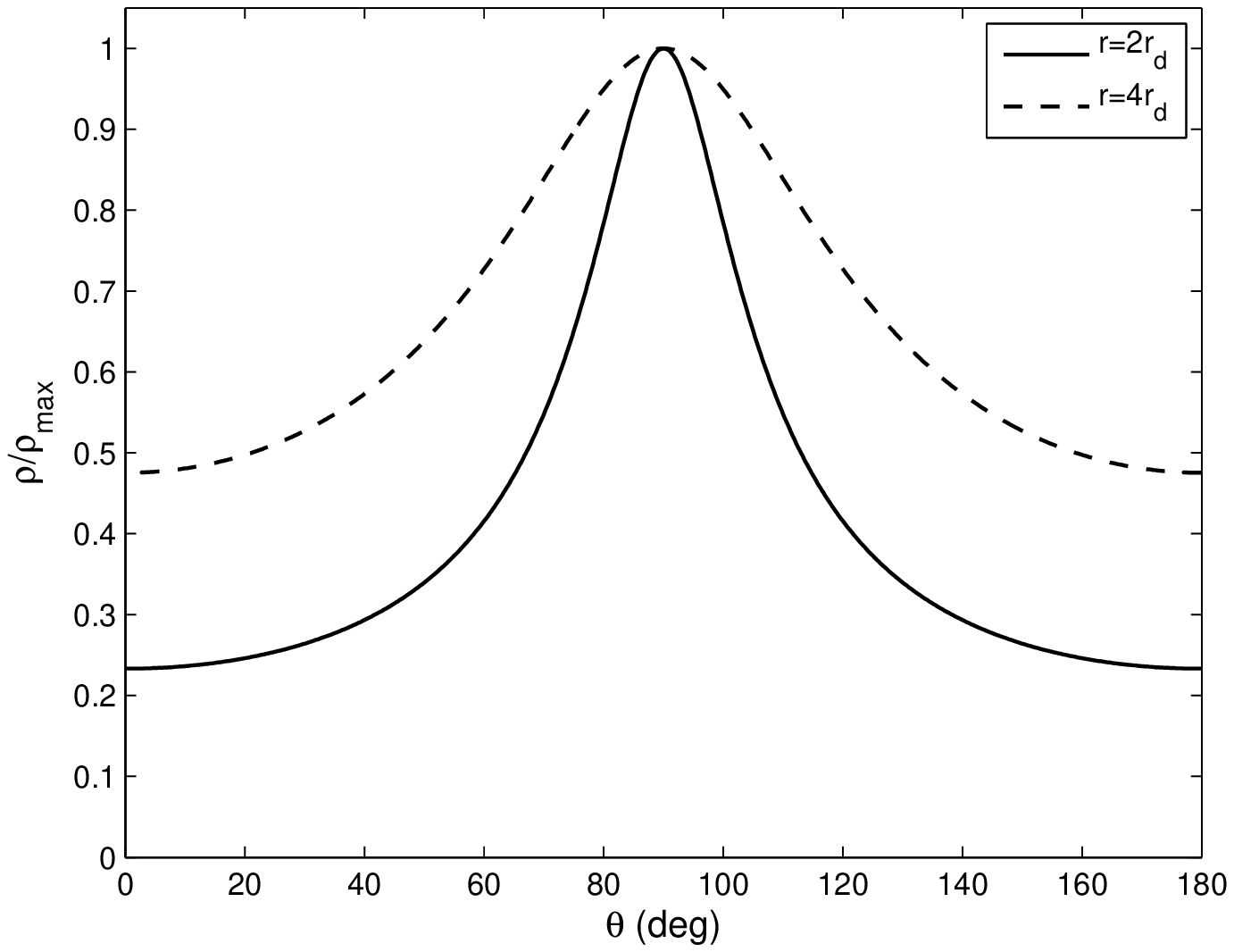}\\
\caption{Density distribution with radius $r$ (left panel) and polar angle
  $\theta$ (right panel). $n_\mathrm{He}=0.1n_\mathrm{H}$ is assumed here. 
  In the left panel, different curves correspond to
  different polar angles, which is $0^\circ$ when along the rotation axis. The thick black
  curve is without any rotation effect for comparison. The right panel shows the density dependence on $\theta$ at $r=2r_d$ and $r=4r_d$.} 
\label{fig:rotation}
\end{center}
\end{figure*}

\citet[]{Shu77} developed the inside-out expansion-wave solution for
the problem of the gravitational collapse of an isothermal
sphere. \citet[]{MP97} applied this solution to the collapse of a
logotropic sphere and also gave the general formulae for a polytropic
sphere. Following their work, we calculate the density profile in the
core. A core with initial density profile of eq. (\ref{den}) is not
stable and a perturbation at the center can trigger collapse of the
innermost region. The position where the material begins to fall
progresses outward, which is called the expansion wave. While the
material collapses inside the expansion wave front, the region outside
it is still hydrostatic. The similarity is always kept during the
whole collapse in this solution.

\citet[]{MP97} did not include the effect of magnetic fields which
can help to support more mass in the core and increase the accretion
rate after collapse starts. We estimate the effect of magnetic fields
from the work of \citet[]{LS97}, who found that the equilibrium
surface density is increased by a factor of $(1+H_0)$ when magnetic
fields are considered, where $H_0$ is a parameter. So the real mass
and density profile should increased by the same factor,
\begin{equation}
M(r,t)=M_{\mathrm{non}}(r,t)(1+H_0)=m(x)(1+H_0)a_t^3t/G,
\end{equation}
and
\begin{equation}
\rho(r,t)=\rho_{\mathrm{non}}(r,t)(1+H_0)=\frac{\alpha(x)(1+H_0)}{4\pi Gt^2},
\end{equation}
where $m(x)$ and $\alpha(x)$ are the similarity variables for mass and
density, and $a_t=[K\gamma_p(4\pi Gt^2)^{1-\gamma_p}]^{1/2}$ has
dimension of velocity. In our case, $H_0$ is set to 1 following the
assumption by MT03.

The collapsed mass at the center now is
\begin{equation}
M(0)=m(0)(1+H_0)a_t^3t/G,
\end{equation}
therefore,
\begin{equation}
M(0)\propto a_t^3t \propto t^{4-3\gamma_p},
\end{equation}
and
\begin{equation}
\dot{M}(0)\propto t^{3-3\gamma_p}\propto t
\end{equation}
in our case, which is consistent with the analysis of
MT03. So, given a certain collapsed mass $M(0,t)$, with the
star formation time (the time that the whole core takes to collapse,
eq. (44) in MT03)
\begin{equation}
t_{*f}=1.29\times 10^5
\Big(\frac{M_\mathrm{core}}{60\;M_\odot}\Big)^{1/4}\Sigma_{\mathrm{cl}}^{-3/4}\;\mathrm{yr}
\end{equation}
and the final collapsed mass $M(0,t=t_{*f})=M_\mathrm{core}=60\;M_\odot$ 
(The whole core collapses at the end, either into the star-disk system or into the outflow 
and escapes), we can
calculate $t$ and further the density $\rho(r,t)$, velocity
$u(r,t)$ and mass $M(r,t)$ at that moment. At some time the
expansion wave will reach the boundary of the core and lead to a backward wave, thus, 
for simplicity  we choose such a
collapsed mass that the expansion wave has not reached the core boundary yet. 
(We will discuss this in detail in Section \ref{sec:star}.) 

\subsubsection{Rotating Infall}
\label{sec:rotating}

We consider a slowly rotating core. For simplicity, we only include
the effect of rotation inside the sonic point, where the infall
becomes supersonic. Once such infall starts, we expect it is difficult
to transfer angular momentum from inside the sonic point to outside
(\citealt[]{TM04}). So we assume that the angular momentum is
conserved inside the sonic point until gas accretes onto the star or
disk.

We use the solution of \citet[]{Ulrich} (referred to below as the
``Ulrich solution'') to describe the velocity field and density profile
inside the sonic point. This solution assumes that a particle with an
initial distance $r_\infty$ from the center, an initial polar angle
$\theta_0$ and angular velocity $\Omega_\infty$ about the axis of
rotation, due to a point mass $M$ at the center, moves in a parabolic
path described by
\begin{equation}
r=\frac{r_d\cos\theta_0\sin^2\theta_0}{\cos\theta_0-\cos\theta}, \label{streamline}
\end{equation}
where
\begin{equation}
r_d=\frac{\Omega_\infty^2r_\infty^4}{GM}=\frac{j_\infty^2}{GM}.
\end{equation}
All particles starting from a spherical shell of radius $r_\infty$
will hit the equatorial plane at radius $r<r_d$, so naturally $r_d$
can be thought of as the radius of the accretion disk.

$r_d$ is also the centrifugal radius which marks the radial extent of
centrifugal balance, inside which the flow will become disk-like
(e.g. \citealt[]{jijina96}). It can be estimated as below:
\begin{equation}
r_d=\frac{6a_g}{5a_i}\beta r=\frac{6a_g}{5a_i}\beta
\Big(\frac{M}{M_\mathrm{core}}\Big)^\frac{1}{3-k_\rho}, \label{rd} 
\end{equation}
where
$a_g=5|E_\mathrm{grav}|r/(3GM^2)=(5/3)(3-k_\rho)/(5-2k_\rho)\rightarrow1.25$,
$a_i=2E_\mathrm{rot}/(Mr^2\Omega^2)=(2/3)(3-k_\rho)/(5-k_\rho)\rightarrow
0.286$, and $\beta =E_\mathrm{rot}/|E_\mathrm{grav}|$ is set to have a
fiducial value of 0.02 for the slowly rotating core, based on
observations of low-mass cores (\citealt[]{Goodman93}). For any
spherical shell at radius $r$ inside the sonic point, it infalls
following eq. (\ref{streamline}), with $r_d$ to be the outer radius of
the disk determined by the position of the sonic point.
\begin{equation}
r_d=\frac{6a_g}{5a_i}\beta
\Big(\frac{M_\mathrm{sp}}{M_\mathrm{core}}\Big)^\frac{1}{3-k_\rho},
\end{equation}
where $M_\mathrm{sp}$ is the mass inside the sonic point. In our
fiducial case, when the collapsed mass is 10.67 $M_\odot$ (8 $M_\odot$
star and 2.67 $M_\odot$ disk, discussed in Section \ref{sec:model}),
the disk radius is 449 AU.

The rotation changes the density distribution from spherical symmetry to
axissymmetry as below
\begin{equation}
\rho(r,\theta)\,\propto\,\Big(1+\frac{\cos\theta}{\cos\theta_0}\Big)^{-1/2}\Big(\frac{\cos\theta}{2\cos\theta_0}+\frac{r_d}{r}\cos^2\theta_0\Big)^{-1}. \label{rotateden}
\end{equation}
This result is valid for an infall rate that is constant with radius,
which is not exactly right for the expansion wave solution of a
polytropic sphere. Thus we only scale the density distribution with
the angular dependence of eq. (\ref{rotateden}) and keep the total
mass in each shell unchanged, so that the infall rate is the same as
the non-rotating case.  Fig. \ref{fig:rotation} shows the radial
density profiles at different polar angles (upper panel) and the
density profiles with polar angle at different radii (lower panel),
for a core with 10.67 $M_\odot$ collapsed at the center. In the upper
panel, the thick black line shows the density profile in a
non-rotating core. The thin curves correspond to different polar
angles. There is a discontinuity at $r\sim 2500$ AU, which marks the
position of the sonic point, only inside of which we consider
rotation. For a real core, this discontinuity would not exist. The
peaks and dips mark the disk radius $r_d$. The lower panel of
Fig. \ref{fig:rotation} shows the dependence of density on $\theta$ at
$2r_d$ and $4r_d$.

\subsection{Accretion Disk}
\label{sec:disk}

Inside the centrifugal radius $r_d$, the infalling flow is
circularized and forms an accretion disk around the protostar. The disk
can provide an efficient way to transfer angular momentum outwards and
enable high accretion rates to form a massive star. The high accretion
rate indicates that the disk is relatively massive (comparable to the
star). But gravitational instability can become very efficient when
the mass of the disk is high enough and lead it to fragment. We assume
the disk mass is always a constant fraction of the stellar mass
(\citealt[]{TM04}):
\begin{equation}
m_{*d}=m_*+m_d=(1+f_d)m_*,
\end{equation}
where $f_d$ is assumed to have a fiducial value of 1/3. Recently,
\citet[]{Kratter10} performed a numerical parameter study on accretion disks
of massive stellar systems and found that a disk can be stable and does not
fragment with an even higher disk-to-star mass ratio ($f_d\sim 1$). 

We follow the description for the disk structure by
\citet[]{Whitney03b}, which is a standard $\alpha$-disk
(\citealt[]{SS73}, \citealt[]{LP74}, \citealt[]{Pringle81},
\citealt[]{Bjorkman97}, \citealt[]{Hartmann98}) with density
distribution
\begin{equation}
\rho=\rho_0 \Big(1-\sqrt\frac{r_*}{r}\Big)
\Big(\frac{r_*}{r}\Big)^\alpha \exp \Big\{-\frac{1}{2}
  \Big(\frac{z}{H(r)}\Big)^2 \Big\}, \label{rho}
\end{equation}
where $r$ is the radial coordinate in the disk midplane, $z$ is the
distance from the midplane, and $r_*$ is the stellar radius. Basically
the density has a power-law profile with radius (smoothed at the
innermost region) and a Gaussian profile with the height. $H$ is the
scale height of the disk,
\begin{equation}
H=H_0\Big(\frac{r}{r_*}\Big)^\beta. \label{height}
\end{equation}
Thus, the disk structure is specified by following parameters: disk
mass $m_d$, disk inner radius $r_{d,\mathrm{in}}$, disk outer radius
$r_d$ (the centrifugal radius), disk scale height $H_0$ at the stellar
surface, and the power indices $\alpha$ and $\beta$.

If the dust is the only source of opacity in the disk, then it equals
to a disk truncated at dust destruction radius $r_\mathrm{sub}$,
inside of which the temperature is too high for dust to
exist. However, in reality, the gas opacity at the innermost region of
the disk can be significant, so the disk should be optically thick
down to the surface of the protostar. Thus, in our fiducial model, the
inner radius of the disk is set to be the stellar radius.

We consider both geometrically thin ($H_0/r_*= 0.01$) and moderately
thick ($H_0/r_*= 0.1$) disks to see the effects of the disk scale
height. $H_0/r_*=0.01$ is the value used by
\citep{Whitney03a,Whitney03b}.  $H_0/r_*=0.1$ is closer to the aspect
ratio expected for a disk composed of dust only and is a typical value
for a moderately thick disk. For these cases, $\alpha$ and $\beta$ are
chosen to be 1.875 and 1.125, respectively, as in the standard
$\alpha$ disk models considered by Whitney. However, in our final
fiducial model, we calculate the disk scale height, $\alpha$ and
$\beta$ self-consistently from the gas and dust opacities, in which
case, fitting with eq. (\ref{rho}) and (\ref{height}) gives
$H_0/r_*=0.06$ and $\alpha=1.75$, $\beta=1.08$.

In the limit of a slowly rotating protostar, the accretion luminosity
from the system is
\begin{equation}
L_{\mathrm{acc}}=\frac{Gm_*\dot{m}}{r_*},\label{totalacc}
\end{equation}
half of which is emitted from the viscous disk (disk accretion
luminosity $L_\mathrm{disk}$) and the other half is emitted when the
material hits the surface of the protostar (hot-spot luminosity
$L_\mathrm{hotspot}$),
\begin{equation}
L_\mathrm{disk}=L_\mathrm{hotspot}=\frac{1}{2}L_\mathrm{acc}.
\end{equation}
For simplicity, we add the hot-spot luminosity to the star's
homogeneously. The star then has a single, enhanced temperature to
describe its black body spectrum. If the disk is truncated at
$r_{d,\mathrm{in}}$, the disk luminosity will be
\begin{equation}
L_{\mathrm{disk}}(r_{d,\mathrm{in}})=\frac{Gm_*\dot{m}}{2r_{d,\mathrm{in}}}\Big(3-2\left(r_*/r_{d,\mathrm{in}}\right)^{1/2}\Big),\label{ldisk}
\end{equation}
which indicates that only when the disk extends to the stellar radius
can the total energy be conserved. In fact, the disk accretion
luminosity is very sensitive to the inner radius of the disk. The
accretion rate $\dot{m}$ of a massive protostar can be very high,
reaching $\sim 10^{-4} - 10^{-3} \;M_\odot$/yr. Here we adopt the
protostar evolution model by MT03, which gives an accretion
rate of $2.40\times 10^{-4}\;M_\odot$/yr for an 8 $M_\odot$ protostar
in a 60 $M_\odot$ core. This accretion rate gives a disk luminosity and a
hotspot luminosity which are comparable to the stellar luminosity.

The disk accretion rate is related to the $\alpha$ parameter of the disk by
\begin{equation}
\dot{m}=\sqrt{18\pi^3}\alpha_\mathrm{disk}V_c\rho_0 H_0^3/r_*
\end{equation}
with $V_c=(Gm_*/r_*)^{1/2}$. Parameters used in our fiducial model
correspond to a case with $\alpha_\mathrm{disk}=1.43$, which is consistent 
with the results of numerical simulations by \citet[]{Krumholz07}, 
in which they found an effective $\alpha_\mathrm{disk}\;\sim1.0-1.6$.

Also, part of the energy may
be used to drive the outflow (mechanical luminosity $L_w$,
MT03). But for now we only
consider the radiative luminosity from the disk and the hot-spot.

\subsection{Outflow Cavity}
\label{sec:outflow}

Powerful bipolar outflows are ubiquitous phenomena around protostars
(e.g. \citealt[]{KP2000}). 
The prevalent interpretation is that outflows are powered by accretion
activity, being driven by spinning magnetic fields that thread the
disk. There are several theoretical models to describe this process
such as the X-wind from the innermost region of the disk
(\citealt[]{Shu94}) and disk wind model (\citealt[]{KP2000}).

A common feature of these models is the production of a bipolar
outflow with momentum distribution $p_w\propto (\sin\theta_w)^{-2}$ for
$\theta_w>\theta_{w0}$, where $\theta_w$ is measured from the outflow axis
and $\theta_{w0}\sim 10^{-2}$ is a small angle (\citealt[]{MM99}). On scales large
compared to the source,
\begin{equation}
\frac{\ud \dot{p}_w}{\ud
  \Omega}=\frac{\dot{p}_w}{4\pi\ln(2/\theta_{w0})(1+\theta_{w0}^2-\cos^2\theta_w)}. \label{outflow}
\end{equation}
We follow the discussion of \citet[]{TM02}. Assuming an opening angle
of $\theta_{w,\mathrm{esc}}$ for the outflow cavity, the outflow material
with $\theta_w>\theta_{w,\mathrm{esc}}$ cannot escape from the core and will
go back to the infalling flow. We parameterize the fraction of the
outflow momentum that escapes from the core with
$f_{w,\mathrm{esc}}$. \citet[]{NS94} showed that the velocity of the
wind is approximately independent of the polar angle so that
$f_{w,\mathrm{esc}}$ can also describe the ratio between the mass loss
rates. We assume that part of the material accreted to the disk will
be transferred to the star at a rate $\dot{m}_*$, and another part will
leave to the outflow at a rate $\dot{m}_w$, while the rest is left in
the disk so the disk grows in a rate $\dot{m}_d$. We also assume that
$\dot{m}_d = f_d \dot{m}_*$, $f_w=\dot{m}_w/\dot{m}_*$ and
$f_{w,\mathrm{esc}}$ are all constant.

Generally, we assume the outflow starts at time $t_1$ when the stellar
mass is $m_*(t_1)$ and the disk mass is $m_d(t_1)$ ($t_1=0$
corresponds to situation where the outflow cavity forms as soon as the
collapse begins.) When $t<t_1$ there is no outflow and we have
\begin{equation}
m_{*d,0}(t_1)=m_{*d}(t_1)=m_*(t_1)+m_d(t_1)=(1+f_d)m_*(t_1),
\end{equation}
where $m_{*d,0}$ is the collapsed mass of the polytropic core 
(a hypothetical star-disk mass if the feedback is absent). After $t_1$ until
the time when the whole core has collapsed $t_f$, we have
\begin{equation}
\dot{m}_*+\dot{m}_d+f_{w,\mathrm{esc}}\dot{m}_w=\cos\theta_{w,\mathrm{esc}}\dot{m}_{*d,0},
\end{equation}
where $\cos\theta_{w,\mathrm{esc}}$ means only part of the infalling core
accretes to the disk because of the existence of the outflow
cavity. Therefore, the instantaneous star formation efficiency is 
\begin{equation}
\epsilon_{*d}\equiv\frac{\dot{m}_{*d}}{\dot{m}_{*d,0}}=\frac{m_{*d}(t_f)-m_{*d}(t_1)}{M_\mathrm{core}-m_{*d,0}(t_1)}=\frac{(1+f_d)\cos\theta_{w,\mathrm{esc}}}{1+f_d+f_wf_{w,\mathrm{esc}}},
\end{equation}
and the mean star formation efficiency is 
\begin{equation}
\bar{\epsilon}_{*d}\equiv\frac{m_{*d}(t_f)}{m_{*d,0}(t_f)}=\frac{m_{*d}(t_f)}{M_\mathrm{core}}
\end{equation}

After the whole core has collapsed, part of material in the disk will accrete
onto the star and rest of them will leave the star-disk system to the outflow
wind, which gives us
\begin{equation}
m_{*f}=m_*(t_f)+\frac{m_d(t_f)}{1+f_w}=\frac{1+f_d+f_w}{(1+f_w)(1+f_d)}m_{*d}(t_f),
\end{equation}
where $m_{*f}$ is the mass of the star finally born out of the core, which is
assumed to be half of the initial total mass of the core $M_\mathrm{core}$, 
i.e, the final star formation efficiency is
\begin{equation}
\epsilon_{*f}\equiv\frac{m_{*f}}{M_\mathrm{core}}=\frac{1+f_d+f_w}{(1+f_w)(1+f_d)}\bar{\epsilon}_{*d}=\frac{1}{2}. \label{epsilon}
\end{equation}

For the models with outflow cavities that we investigate in this
paper, we make the assumption that the outflow cavity has only now
formed when $m_*(t_1)=8\;M_\odot$, so $m_{*d}(t_1)=10.67\;M_\odot$. The
validity of this assumption will be examined in a future paper.
Combining eq. (\ref{outflow}) to (\ref{epsilon}) we solve for the
opening angle $\theta_{w,\mathrm{esc}}$ and $f_{w,\mathrm{esc}}$
simultaneously.  For $\epsilon_{*f}=0.5$ and $f_d=1/3$, we find that
when $f_w=0.1\sim 0.8$, the opening angle varies from 64$^\circ$ to
45$^\circ$ and $f_{w,\mathrm{esc}}$ varies from 0.91 to 0.84. So the
fraction of the outflow coming back to the infalling core is always
small, and the opening angle is typically large except when $f_w$ is
close to 1. We choose $f_w=0.6$ as our fiducial value, making
$f_{w,\mathrm{esc}}=0.86$ and the opening angle of the outflow cavity
to be 51$^\circ$. In our models, the cavity wall follows the
streamline of the rotating infalling material
(eq. \ref{streamline}). For now we set the outflow cavity to be empty
(i.e. the optically thin limiting case). One would expect it to be
relatively free of dust if most of the outflow is launched from the
region of the disk inside the dust destruction front and assuming
there is little time for new dust grains to form in the rapidly
expanding outflow. We will improve upon this assumption by studying
the detailed density distribution in the outflow cavity in our
next paper. Since the cavity does not change the density distribution
in rest of envelope, the total mass of the material in the envelope
after outflow cavities are carved out is about half its original
value.

\begin{figure*}
\begin{center}
\includegraphics[width=0.9\columnwidth]{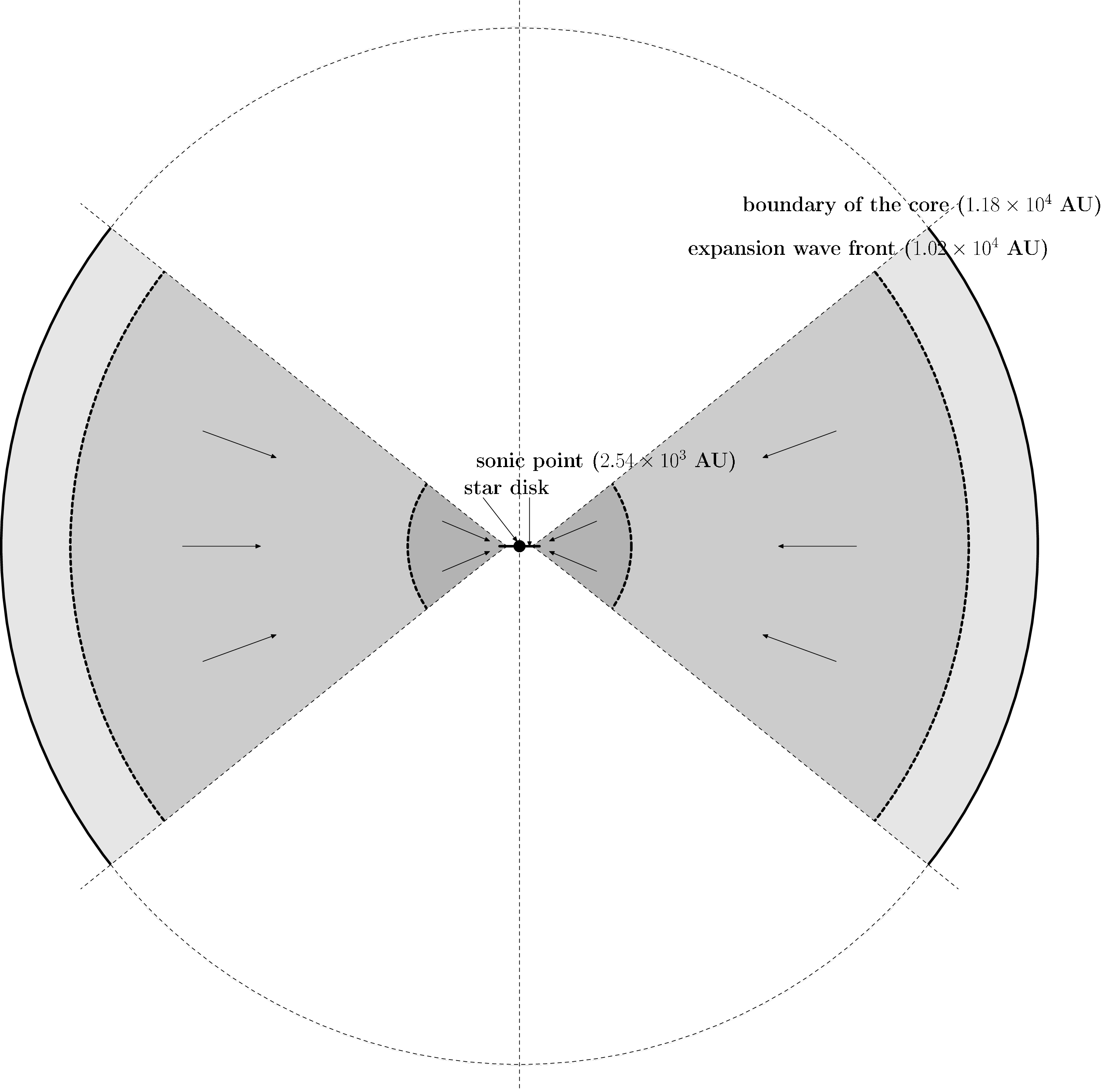}
\includegraphics[width=0.9\columnwidth]{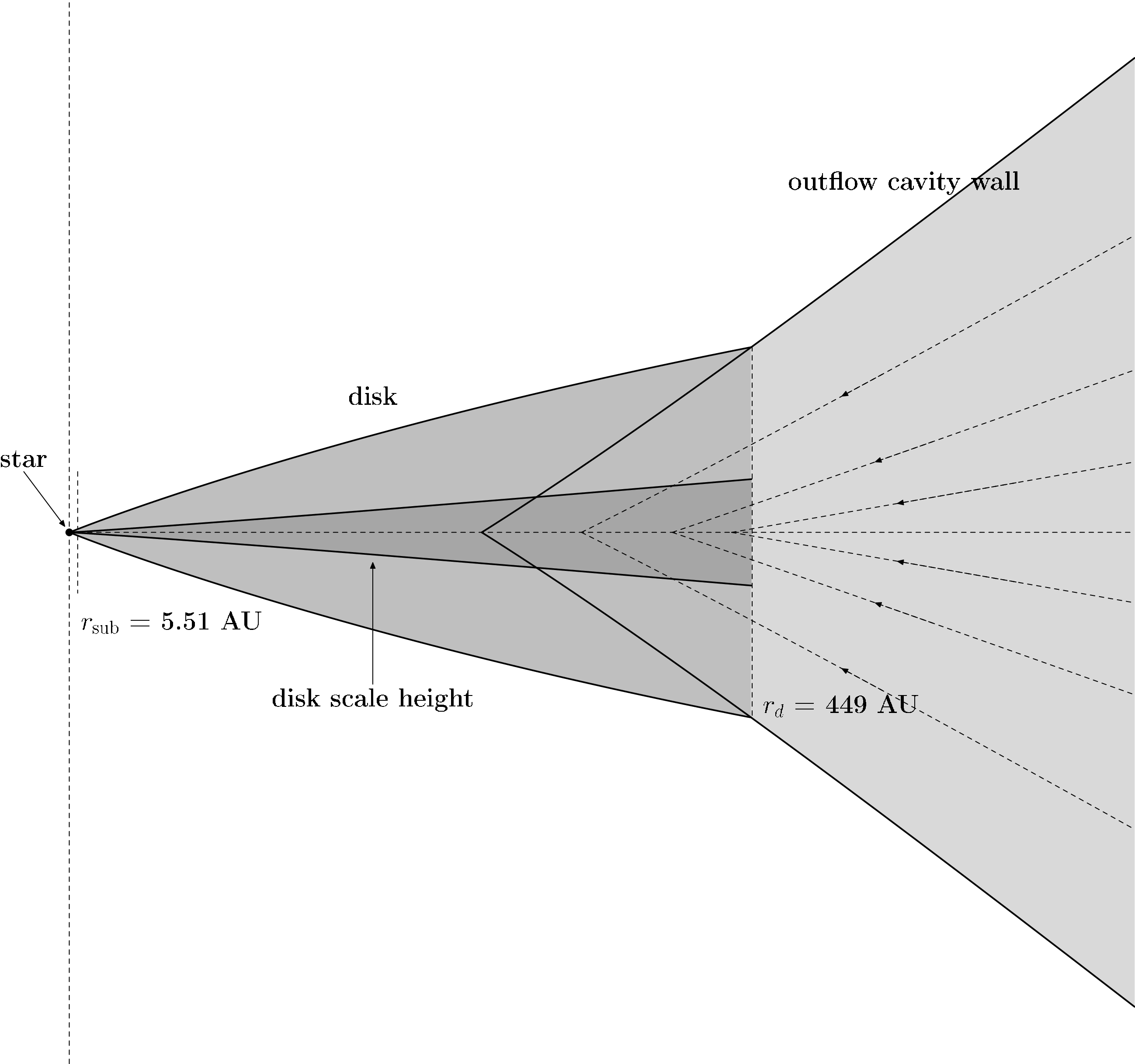}\\
\caption{Structure of the massive star forming core in the fiducial
  model. The right panel is a zoom-in view of the central region of
  the left panel. The positions of core boundary, expansion wave
  front, sonic point (inside which we consider rotation), disk scale
  height, dust destruction radius and the outer radius of the disk are
  marked. The outer boundary of the disk is chosen to be a surface
  with constant density joining the outflow cavity wall at $r_d$.}
\label{fig:sketch}
\end{center}
\end{figure*}

\subsection{Protostar}
\label{sec:star}

Assuming that the core mass is a constant, the mass of the central
protostar indicates its age and evolutionary stage. At some moment,
the outgoing expansion wave front will reach the boundary of the core
and, depending on the properties of the boundary, may lead to a
backward wave and a breakdown of self-similarity. In our case, this
happens when the collapsed mass reaches $m_{*d,0}\sim 12\; M_\odot$.
For the present paper we have chosen to consider a series of models
with $m_*=8\;M_\odot$, and thus a maximum central collapsed mass of
10.67 $M_\odot$ for those cases with rotating infall to a disk. Thus
the central object is on the verge of becoming a ``massive
protostar'', following the definitions of \citet[]{Beuther07} and
\citet[]{Tan08}. The full evolutionary sequence, including a treatment for
greater masses, will be considered in a future paper.

MT03 studied the evolution of massive protostars, and for
$m_*=8\;M_\odot$ and $\Sigma_{\rm cl}=1\:{\rm g\:cm^{-2}}$, the
protostar is not expected to have yet contracted to the main sequence.
They calculated the values of radius and luminosity of the protostar:
$r_*=12.05\;R_\odot$ and $L_*=2.81 \times 10^3\;L_\odot$. For simplicity,
we assume a black body spectrum for the star, with surface temperature
$T_*=1.22\times 10^4$ K in this condition. For models with an active
disk, we also add the hot-spot luminosity to the stellar spectrum,
assuming it is emitted homogeneously from the stellar surface, which
means the temperature now is $T_{*,\mathrm{hotspot}}=1.43\times
10^4$ K. 

For the cases with $\Sigma_{\rm cl}$ = 0.316 g cm$^{-2}$, we use the
following parameters: $r_*=11.3\;R_\odot$, $T_*=1.25\times10^4$ K, and
$T_{*,\mathrm{hotspot}}=1.37\times10^4$ K. For the cases with
$\Sigma_{\rm cl}$ = 3.16 g cm$^{-2}$, these parameters are:
$r_*=5.93\;R_\odot$, $T_*=1.74\times10^4$ K, and
$T_{*,\mathrm{hotspot}}=2.62\times10^4$ K.

\subsection{Model Series}
\label{sec:model}

\begin{table*}
\begin{center}
\caption{Properties of features included in the model series.\label{table1}}
\begin{tabular}{c|c|c|c|c}
\hline
\bf Models & \bf Star & \bf Disk & \bf Envelope & \bf Outflow \\
\hline
1 & 8$M_\odot$ & no & 52$M_\odot$, $\rho\propto r^{-1.5}$ & no \\
\hline
2 & 8$M_\odot$ & no & 52$M_\odot$, expansion wave & no\\
\hline
\multirow{2}{*}{3} & \multirow{2}{*}{8$M_\odot$} & $1/3m_*$, thin ($H_0/r_* = 0.01$), &
$49.333M_\odot$, expansion & \multirow{2}{*}{no} \\
& & passive, $r_{d,\mathrm{in}}=r_\mathrm{sub}$ & wave, rotation & \\
\hline
\multirow{2}{*}{4} & \multirow{2}{*}{8$M_\odot$} & \multirow{2}{*}{as above} &
only $\sim 29 M_\odot$ left, & \multirow{2}{*}{yes} \\
& & & expansion wave, rotation & \\
\hline
\multirow{2}{*}{5} & \multirow{2}{*}{8$M_\odot$} & $1/3m_*$, thin, & \multirow{2}{*}{ as above
 } & \multirow{2}{*}{yes} \\
& & $r_{d,\mathrm{in}}=r_\mathrm{sub}$, active & & \\
\hline
\multirow{2}{*}{6} & \multirow{2}{*}{8$M_\odot$} & $1/3m_*$, active, $r_{d,\mathrm{in}}=r_\mathrm{sub}$& \multirow{2}{*}{ as above
 } & \multirow{2}{*}{yes} \\
& & thick ($H_0/r_* = 0.1$) & & \\
\hline
\multirow{2}{*}{7} & \multirow{2}{*}{8$M_\odot$} & same as Model 8, & \multirow{2}{*}{ as above
 } & \multirow{2}{*}{yes} \\
& & except $r_{d,\mathrm{in}}=r_\mathrm{sub}$  & & \\
\hline
8 & \multirow{2}{*}{8$M_\odot$} & 1/3$m_*$, active, & \multirow{2}{*}{as above} & \multirow{2}{*}{yes}\\
(fiducial) & & $H_0/r_*=0.06$, $r_{d,\mathrm{in}}=r_*$ && \\
\hline
\end{tabular}
\end{center}
\end{table*}

\begin{table*}
\begin{center}
\caption{Parameters of the models comparing different column density. \label{table2}}
\begin{tabular}{l|c|c|c}
\hline
& Model 8 (fiducial) & Model 8l & Model 8h\\
\hline
mean surface density $\Sigma_\mathrm{cl}$ (g/cm$^2$) & 1 & 0.316 & 3.16\\
\hline
core radius $R_\mathrm{core}$ (pc) & 0.057 & 0.10 & 0.032\\
\hline
position of expansion wave front $r_\mathrm{ew}$ (pc) & 0.049 & 0.088 & 0.028\\
\hline
formation time $t_f$ (yr) & $1.29 \times 10^5$ & $3.06 \times 10^5$ & $5.44 \times 10^4$ \\
\hline
position of sonic point $r_\mathrm{sp}$ (AU) & $2.57\times 10^3$ & $4.57\times 10^3$ & $1.45\times 10^3$\\
\hline
outer radius of disk $r_d$ (AU) &  449.4 & 801.4 & 253.4\\
\hline
disk accretion rate $\dot{m}$ ($M_\odot$/yr) & $2.398\times 10^{-4}$ & $1.035\times 10^{-4}$ & $5.667\times 10^{-4}$ \\
\hline
stellar radius $r_*$ ($R_\odot$) & 12.0 & 11.3 & 5.93\\
\hline
stellar surface temperature $T_*$ (K) & $1.22\times 10^4$ & $1.25\times 10^4$ & $1.74\times 10^4$\\
\hline
stellar + hotspot temperature $T_{*,\mathrm{hotspot}}$ (K) & $1.42\times 10^4$ & $1.37\times 10^4$ & $2.62\times 10^4$\\
\hline
stellar luminosity $L_*$ ($L_\odot$) & $2.81\times 10^3$ & $2.82 \times 10^3$ & $2.84\times 10^3$\\
\hline
total accretion luminosity $L_\mathrm{acc}$ ( $L_\odot$) & $4.90\times 10^3$ & $2.25\times 10^3$ & $2.36\times 10^4$ \\
\hline
disk scale height at stellar surface $H_0/r_*$ & 0.06 & 0.053 & 0.078\\
\hline
\end{tabular}
\end{center}
\end{table*}

In order to study the effects of all the features discussed above on
SEDs and images, we construct a model series starting from the
simplest to the one containing all these features. All the models
assume an initial core mass of 60 $M_\odot$ from which an 8 $M_\odot$
protostar has formed at the center.
 
In Model 1, we assume a spherical symmetric density distribution in
the core with a power-law dependence on the radius, $\rho\propto
r^{k_\rho}$. The total mass in the envelope is $52\;M_\odot$ 
since an $8 M_\odot$ protostar has already formed at the center.

In Model 2, We change the radial density distribution in the envelope
to the expansion wave solution (similar to the thick line in the upper
panel of Fig. \ref{fig:rotation}, but here the expansion wave front is at
$r=0.0408$ pc and the envelope mass is again fixed at $52\:M_\odot$). 

We begin to consider rotating infall inside the sonic point and thus a
disk around the star in Model 3. The disk is geometrically thin
($H_0/r_* = 0.01$) and passive (no accretion luminosity) with the
inner radius set to be the dust destruction radius $r_\mathrm{sub}$,
 which is empirically determined to be (\citealt[]{Whitney04})
\begin{equation}
r_\mathrm{sub}=r_*(T_\mathrm{sub}/T_*)^{-2.1}
\end{equation}
where $T_\mathrm{sub}$ is the dust sublimation temperature and we adopt
$T_\mathrm{sub}=1600$ K.
The expansion wave
front now reaches $r=0.0494$ pc for the collapsed mass now is
$m_*+m_d=10.67\;M_\odot$. The envelope mass is now $49.33\;M_\odot$.

Outflow cavities are added in from Model 4. We keep the density
profile in the envelope unchanged, which corresponds to a case that
the outflow has just swept up the material to form the bipolar
cavities. The envelope mass is now $\sim 29\:M_\odot$.

The accretion luminosities (both from disk and hot-spot) are turned on
in Model 5. However, since the disk is truncated at the dust
destruction radius which in this case is $\sim 98.4\; r_*$, most of the
disk accretion luminosity is lost and the rest of it 
($L_\mathrm{disk}=25.29\;L_\odot$) is much lower than
the hotspot luminosity ($L_\mathrm{hotspot}=2.45\times10^3\;L_\odot$). 
Note that in Robitaille model,
part of this missing disk luminosity is included
in the stellar $+$ hotspot luminosity.

In Model 6 we adjust the disk to be a geometrically thick one
($H_0/r_* = 0.1$) to see the effects of the height of the disk.

In our fiducial model, the disk is extended to the stellar radius so
that the accretion luminosity can be conserved. Inside the dust
destruction front ($T >1600$ K), gas opacities are used. Here we assume
that the $\alpha$-disk model is still valid.
The scale height, $H_0$, and radial and vertical scaling parameters
$\alpha$ and $\beta$ are calculated self-consistently from the assumed
opacities and other stellar and envelope parameters: $H_0/r_*=0.06$,
$\alpha=1.75$, and $\beta=1.08$.  Fig. \ref{fig:sketch} is a schematic
of the structure in the fiducial model. The positions of core
boundary, expansion wave front, sonic point, outer radius of the disk,
dust destruction radius, and the scale height of the disk are all
marked. Unlike the previous models, in which disk material can exist
at any height (though the density is very low at large $z$), we
truncate the disk at a fixed density of $2\times10^{-15}\;\mathrm{g}\;\mathrm{cm}^{-3}$ 
so that the disk atmosphere joins smoothly
with the infalling envelope. (The input density profile can be seen in
the lower panel of Fig. \ref{fig:tdinit}.) We expect that the
existence of optically thick gas in the innermost region of the disk
can make a significant difference. In order to investigate this
effect, we construct Model 7 to be the same as the fiducial model
except without gas disk ($r_{d,\mathrm{in}}=r_\mathrm{sub}$). And the
fiducial model is labeled with Model 8.  Table \ref{table1} lists the
differences of all these models.

The column density of the clump where the core is embedded in can
affect the surface pressure of the core, therefore its size, outer
radius of the disk, accretion rate and also the scale height of the
disk. In order to investigate this effect, we consider other two
values for the mean surface density of the clump,
$\Sigma_\mathrm{cl}=0.316$ and 3.16 g/cm$^2$. These correspond to a
change in column density of the core by a factor of 10 and the surface
pressure by a factor of 100. These two models are labeled with Model
8l (low surface density) and 8h (high surface density). Table
\ref{table2} lists the differences of these three models.

\section{Simulations}
\label{sec:simulation}

\begin{figure*}
\begin{center}
\includegraphics[width=0.8\textwidth]{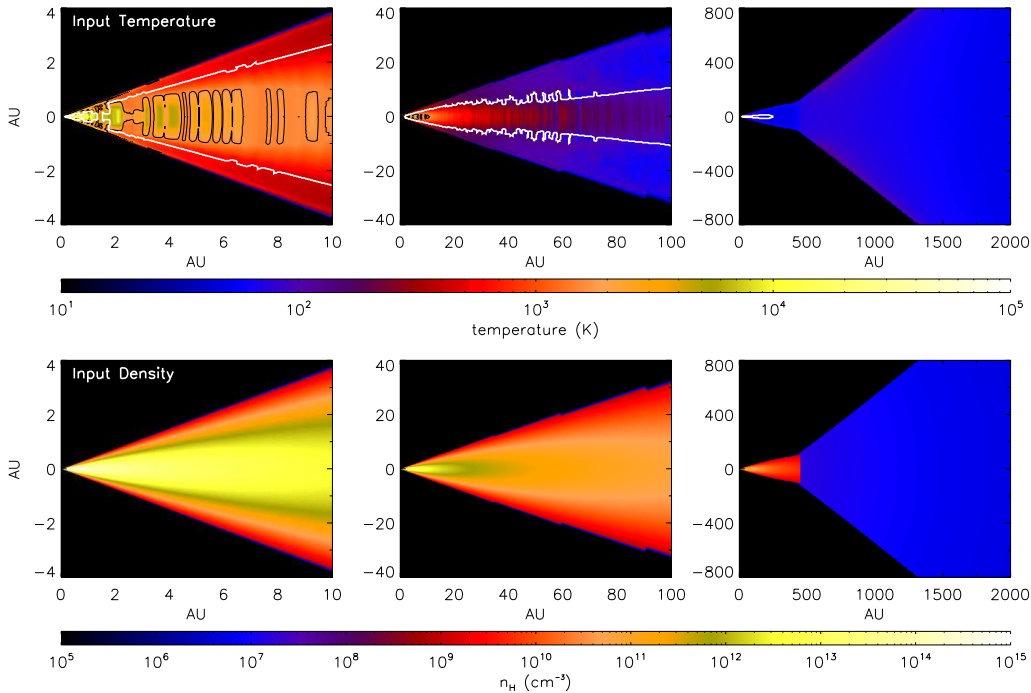}\\
\caption{Input temperature and density for our fiducial model. 
$n_\mathrm{He}=0.1n_\mathrm{H}$ is assumed here. The
  black contours in the upper panels correspond to the dust destruction
  front (1600K), cells inside which are assigned with gas
  opacities. The white contours show the photon-diffusive region calculated in the fiducial model, 
  a photon emitted inside which will move to the surface through a path with constant $r$ and emitted again from the surface. The temperatures of the cells on that path are calculated with grey atmosphere approximation, which is why the the dust destruction fronts have these vertical structures. Also, since the frequency of the photon, when it finally leave the photon-diffusive region, is calculated from the surface temperature, these vertical structures inside it should not affect the results.}
\label{fig:tdinit}
\end{center}
\end{figure*}

\begin{figure}
\begin{center}
\includegraphics[width=\columnwidth]{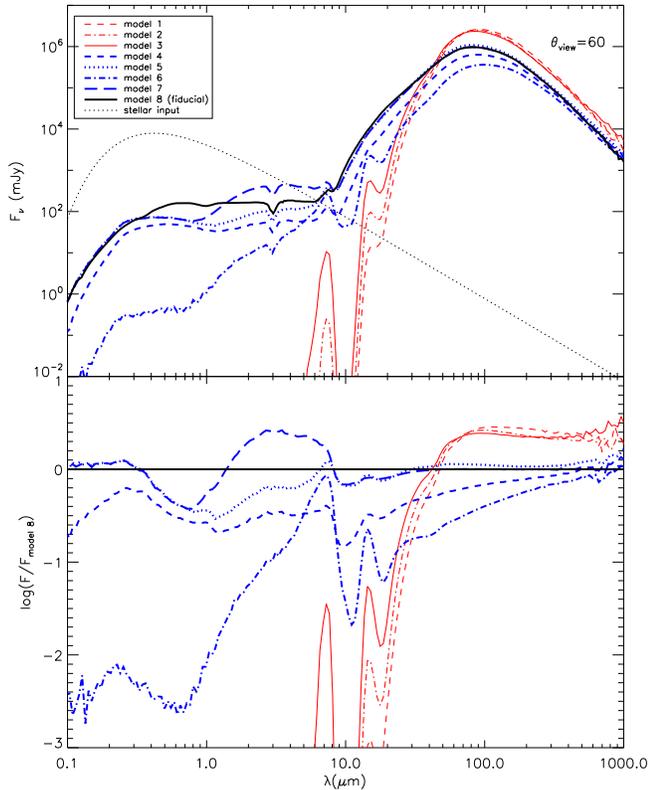}\\
\caption{SEDs for the model series, assuming a distance of 1 kpc. A
  typical inclination of $60^\circ$ is chosen. The upper panel shows
  the SEDs and the lower panel shows the differences of each model
  from the fiducial one. The black solid curve is for the fiducial model,
  the red thin curves are for Model 1 - 3 (models without outflow
  cavities), and the blue thick curves are for Models 4 - 7 (models with
  outflow cavities) The black dotted line is the input stellar
  spectrum (black-body) which is used for all the models.}
\label{fig:sed_series_60}
\end{center}
\end{figure}

\begin{figure}
\begin{center}
\includegraphics[width=\columnwidth]{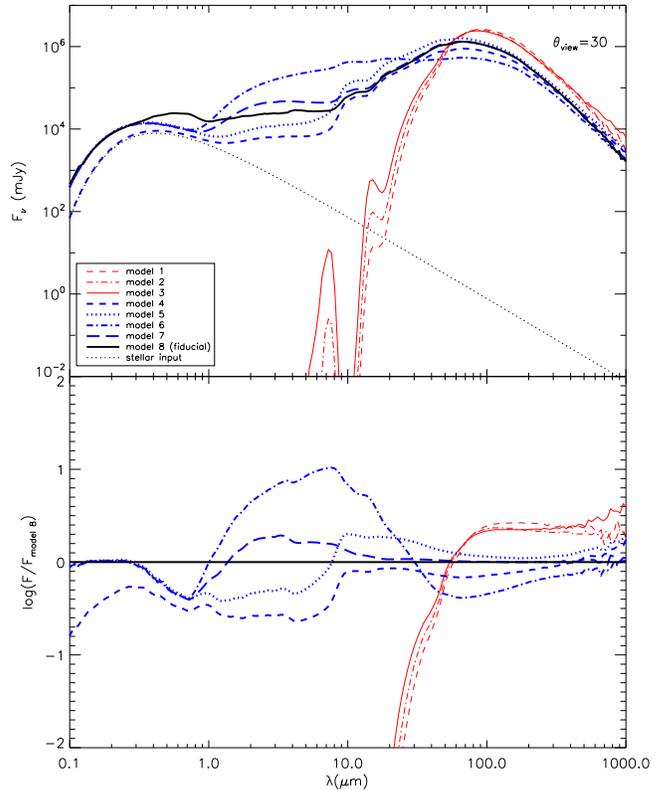}\\
\caption{The same as Fig. \ref{fig:sed_series_60}, except for an inclination of $30^\circ$.}
\label{fig:sed_series_30}
\end{center}
\end{figure}

\begin{figure*}
\begin{center}
\includegraphics[width=0.8\textwidth]{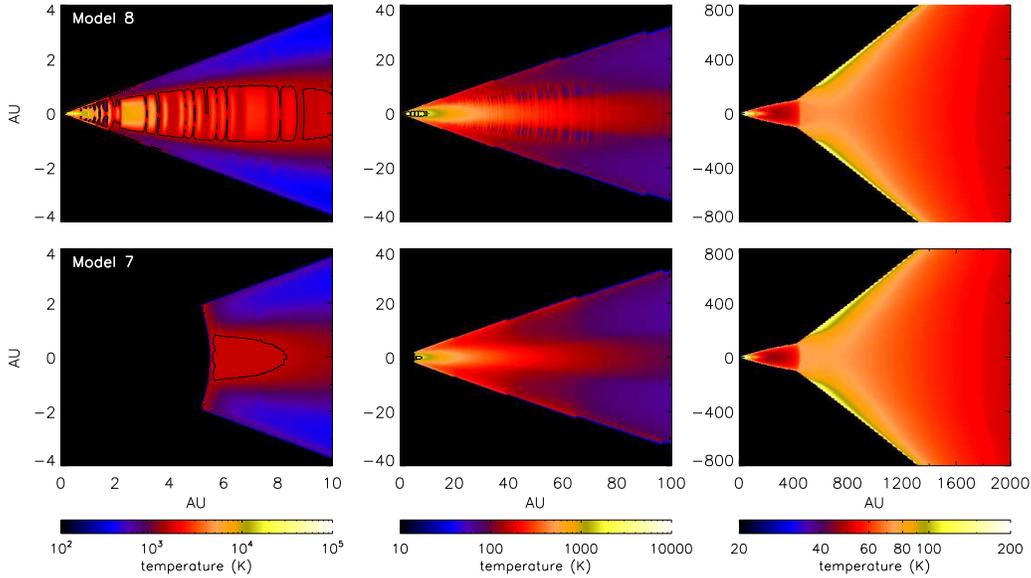}\\
\caption{Output temperature distribution of Model 7 and 8, showing the
  difference the gaseous innermost disk makes. The black contours are
  dust destruction front ($T$=1600K). 
}
\label{fig:tdiffus_78}
\end{center}
\end{figure*}

\begin{figure}
\begin{center}
\includegraphics[width=\columnwidth]{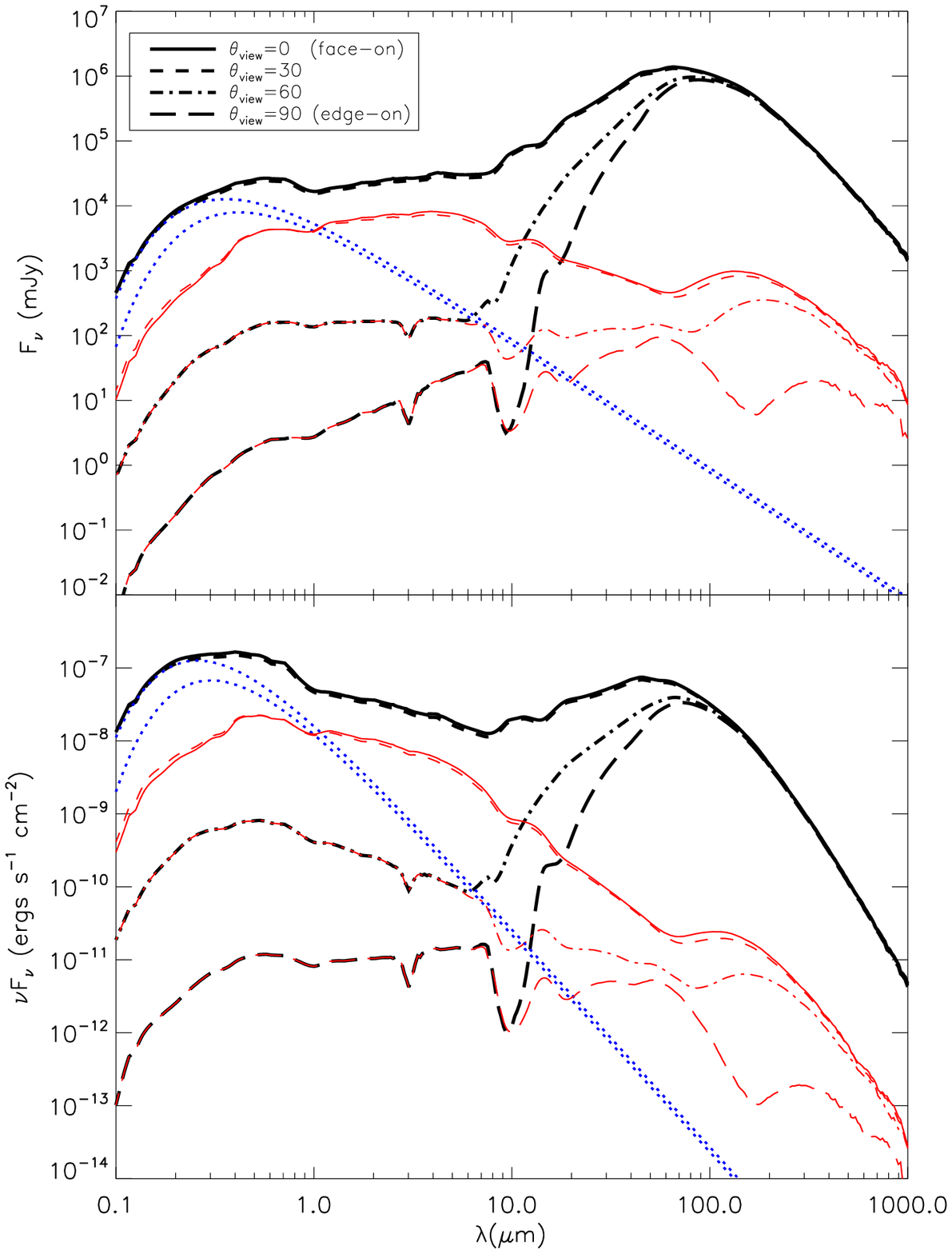}\\
\caption{SEDs of the fiducial model (Model 8) at 4 inclinations,
  assuming a distance of 1 kpc ($F_\nu$ in upper panel and $\nu F_\nu$ in lower panel).
   The thicker lines are total fluxes
  and the thiner red lines are only the scattered light. The lower dotted blue
  line is the input stellar spectrum (black-body). The upper dotted
  line is the total luminosity of star and hot-spot emitted as a
  black-body.}
\label{fig:sed_fiducial}
\end{center}
\end{figure}

We use the Monte Carlo radiation transfer code by
\citet[]{Whitney03b} to perform our calculations. This code includes
thermal emission, non-isotropic scattering and polarization due to
scattering from the dust in a spherical-polar grid, using the method
of Monte Carlo radiative equilibrium by \citet[]{BW01}. This method
requires that the opacities are set up at the beginning of each run
and kept unchanged. During the run, the temperature in each cell is
always being corrected by calculating the balance of absorption and
emission of new photon packets.  This becomes a problem when gas
opacities are included because they are highly dependent on the
temperature and density. So we have to iterate to obtain the correct
temperature profile for the disk (especially for the inner region) and
then use it to set up opacities for each cell.
   
We choose the analytical temperature profile from $\alpha$-disk model
in our case as an initial condition.  After several iterations, the
temperature of the outer disk $\gtrsim$ 4 AU becomes quite
stable. However, because of the discontinuity of the opacities at the
transition region between gas and dust, the temperature profile in the
inner region oscillates between two phases. Even though the difference
can be large between results from two adjacent steps
(i.e. $T_n(r,\theta)$ and $T_{n-1}(r, \theta)$), the averaged
temperature profile of two adjacent step should keep similar
(i.e. $\log(T'_n(r,\theta))=(\log(T_n(r,\theta))+\log(T_{n-1}(r,\theta)))/2$).
So we stop the iteration when $T'_n\sim T'_{n-1}$, more specifically,
when standard deviation of the distribution of $({\rm log} T'_n - {\rm
  log} T'_{n-1})/{\rm log} T'_n$ for the cells inside 4 AU becomes
consistently $<0.1$, which corresponds to a variation of
$\sim$ 25\%. Then we use the temperature profile which is higher in the
midplane to be the input for the final run with large number of
photons to generate SEDs and images. We performed simulations with
both input temperature profiles and saw no significant differences on the SEDs
and images. We also doubled the photon number for the iteration but
found no significant dependence of the standard deviation on the
photon number. It should be noted here that 0.1 standard deviation
inside 4 AU is only an arbitrary standard.  
The upper panel of Fig. \ref{fig:tdinit} shows the input temperature
profile. The black contours correspond to the dust destruction front
($T=1600$ K), inside which gas opacities are assigned depending on the
temperature and density. We can also see that the dust destruction
front extends to $\sim 10$ AU in the midplane and $\sim 3$ AU on the
surface of the disk, which agrees with the estimate of the dust
destruction radius of 5.5 AU in Model 3 - 7. 
We note that temperature iteration is only used in Model 8, while in
other models, the dust opacities are assigned only depending on the
region and the density. Also, this input temperature is only used to
assign opacities. It is not the initial temperature condition for each
run.

The inner region of the disk around the midplane is very optically
thick and so detailed radiative transfer simulation becomes very
time-consuming. Thus in the code the grey atmosphere approximation is
used to describe this region (photon-diffusive region). A photon
generated inside this photon-diffusive region will move to the surface
of the region by following a path with same $r$, and is then emitted
with a frequency calculated based on the temperature of the surface
cell. The temperatures of the cells on that path are calculated
accordingly with the grey atmosphere approximation. In models except
the fiducial one (Model 3 - 7), as in the original code, a cell is set
to be in the photon-diffusive region if the optical depth from
$z=\infty$ to it is larger than 10. In the fiducial model, we change
this to a more restrictive local definition, that the photon-diffusive
region is where the mean free path is smaller than 0.1 $H(r)$. Note
that as long as this photon-diffusive region is set small enough, it
should not affect the final results.

A 3000 $\times$ 1499 grid is used to resolve the $r \times \theta$
space. In $r$ space, $\sim600$ cells are used to resolve the disk with
a finer grid in inner region, $\sim150$ cells are used to cover the
region inside $\sim 6$ AU in the fiducial model. In $\theta$ space,
the grid is finer in the disk (especially around the midplane) and
around the opening angle of the outflow cavity
$\theta_{w,\mathrm{esc}}$. $\sim 700$ cells are used between
$20^\circ$ above and below the disk midplane and $\sim 300$ cells are
used between $50^\circ$ and $53^\circ$
($\theta_{w,\mathrm{esc}}=51^\circ)$.

For each run, SEDs at ten inclinations (evenly distributed in cosine
space) can be produced simultaneously, while if the ``peeling-off''
mode is used, images and SEDs with higher signal-to-noise ratio are
produced for the particular ``peeling-off'' angle. The ``peeling-off''
mode is very time-consuming. For most of our models, $10^8$ photons
are used in one run, and it typically takes several days to one week
running on a single processor. This number of photon packets is still
not perfect for an image of the whole core, especially for those at
wavelengths with low fluxes. Because this code does not enable
parallel computing now, in order to save time, for each model we run
10 times with different random seeds simultaneously on different
processors, and superposed their results, making our final images
contain $10^9$ photons.

Images are produce at several wavelengths and convolved with filter
functions for comparison to observations. The code has already
included filters such as Spitzer IRAC filters at 3.6, 4.5, 5.8 and 8.0
{\micron}, Spitzer MIPS filters at 24, 70 and 160 {\micron}, and 2MASS
J, H, K bands. We also add in the filters of GTC-CanariCam, Herschel
PACS and SPIRE,
and SOFIA FORCAST. We will show images both before and after
convolution with resolution of these particular instruments.

\subsection{Dust Opacity}
\label{sec:dust}

We use three dust grain models for different regions: (1) the
envelope; (2) lower density regions of the disk, and (3) the densest
regions in the disk ($n_{\mathrm{H}_2}>10^{10}\;\mathrm{cm}^{-3}$, the
criterion used by \citealt[]{Robitaille06}). For our present models
there is no dust (or gas) in the outflow cavities. The default dust
models in the code are used without any change
(\citealt[]{Whitney03a}).

The dust model used in the envelope is based on that derived by
\citet[]{KMH94} for the diffuse interstellar medium (ISM). The size
distribution is modified by \citet[]{Whitney03a} to fit an extinction
curve typical of the more dense regions of the Taurus molecular cloud
with $R_V=4$. These grains also include a water ice mantle covering
5\% of the radius. For the lower density regions of the disk, the
grain model of \citet[]{Cotera01} is used. It has grains larger than
ISM grains, but not as large as the disk midplane grains.

For the densest regions of the disk, we use the dust model with large
grains ($\sim$1 mm; model 1 in \citealt[]{Wood02}), which fit the
HH 30 disk SED well. Compared to ISM grains, the larger dust grain
model has a shallower wavelength-dependent opacity: lower at short
wavelengths and larger at long wavelengths.

\subsection{Gas Opacity}
\label{sec:gas}

For most regions of the disk, dust grains dominate the opacity, even if
the mass ratio between dust and gas is low (0.01 is used in our
models). However, in the innermost region of the disk, where the dust
cannot exist ($T>1600$ K), opacity is dominated by gas. Especially when
the temperature is high ($\sim 10^4$ K), the mean opacities of the gas
can be comparable or higher than dust opacities. As discussed in
Section \ref{sec:disk}, most of the disk accretion luminosity is from
this innermost region. Thus, gas opacity should be included in a
realistic model.

In order to simulate the frequency of each photon packet emitted from
the gas-dominated region correctly, not only Rosseland or Planck mean
opacities, but also the frequency-dependent opacities are
needed. Besides, the gas opacity is highly dependent on the
temperature and the density, which make our problem much more
complicated. Since our aim is only to simulate the disk emission
correctly, especially in near and mid-IR, rather than to simulate the
details of line profiles, we smooth the monochromatic opacities for
simplicity and smaller memory demand. Also, we assign opacities to a
grid in temperature and density, rather than to interpolate to obtain
opacities at exact temperature and density.

For temperatures higher than $\sim 3000$ K, we adopt gas opacities from
OP project (\citealt[]{Seaton94}, \citealt[]{Badnell05} and
\citealt[]{Seaton05}). They provide monochromatic gas opacities of
hydrogen, helium and other 15 elements, in large ranges of temperature
($\sim 3000$ K to $10^8$ K, we only use those of temperature up to
$\sim 10^6$ K in our present model since the maximum temperature 
we can reach in the models is less than this) and density.  
The opacities are mainly due to the line absorption, ionization, H$^-$
bound-free absorption, electron-hydrogen/helium free-free
absorption. For scattering, Thompson scattering with a collective
effect is considered (\citealt[]{Boercker87}).
For regions with $T<3000$ K, the opacity model of \citet[]{AF94} is
adopted. However, they do not provide monochromatic opacities for a
range of temperatures and densities. So we use the monochromatic
opacities shown in Fig. 4 of \citet[]{AF94} which is for a temperature
of 2000 K and a density of $8\times 10^{-12}\; \mathrm{g/cm}^3$, and
rescale it for other temperatures and densities based on their
Rosseland mean opacities. Fortunately, the temperature range for this
model is quite narrow (1600K - 3000K). Alexander and Ferguson's model
includes both gas and molecules. At $\sim 2000$K, the total opacity is
dominated by molecules, especially H$_2$O and TiO. Atomic lines and CO
lines are important at some wavelengths. Other continuous sources
include H$^-$ absorption and Rayleigh scattering of H and H$_2$.

In this way we construct an opacity grid of temperature and
density. The temperature varies from 1600 K to 10$^6$ K with an
interval of 0.1 in logarithmic space, and the density varies from
$\sim 10^{-8}$ to $\sim 10^{-15}$ g/cm$^3$ with an interval of 0.5 in
logarithmic space. For each $T$ and $\rho$, a monochromatic opacity
file is assigned, so that we have totally 148 gas opacity files in our
present model. At the beginning of each run, a temperature profile is
read in, and in each cell the gas opacity file with closest $T$ and
$\rho$ to the read-in values is chosen. The opacities are not changed
during the calculation.

Since here we are not concerned with line absorption and emission, it
is better to smooth the monochromatic opacities to save computing
memory. In the code, three important values are calculated: 1)
Rosseland mean opacity 
\begin{equation}
\frac{1}{\kappa_\mathrm{R}}=\frac{\int_0^\infty
(\partial B_\nu / \partial T) / \kappa_\mathrm{ext}(\nu)
\ud \nu}{\int_0^\infty (\partial B_\nu / \partial T) \ud \nu},
\end{equation}
used to determine the photon-diffusive region inside the disk; 2) Planck
mean opacity 
\begin{equation}
\kappa_\mathrm{P}=\frac{\int_0^\infty \kappa_\mathrm{abs}(\nu)
B_\nu \ud \nu}{ \int_0^\infty B_\nu \ud \nu},
\end{equation}
used to calculate the energy equilibrium in a cell; and 3) 
\begin{equation}
P_\nu=\int_0^\nu
\frac{\kappa_\mathrm{abs}(\nu)}{ \kappa_\mathrm{P}} \frac{\partial B_\nu}{\partial T} \ud \nu,
\end{equation} 
where $P_\nu$ is the probability that a photon packet is emitted from a cell with a frequency 
between 0 and $\nu$ (\citealt[]{Bjorkman97}). 
For $\kappa_P$ and $P_\nu$ it is best to smooth opacities
with linear averaging, thus giving better estimates of equilibrium
temperatures and photon frequencies. However, this method tends to
increase $\kappa_R$ by $\sim$10\% or more. Averaging ${\rm log}\kappa$
reduces the importance of line absorption and yields more accurate
Rosseland mean opacities. For our problem, the precise location of the
boundary of the photon-diffusive region is is not very important, so
we smooth opacities with linear averages. The original gas opacity files contain $10^4$ frequencies between $h\nu/kT=0.001$ and 20. We smooth them by averaging 50 adjacent frequencies. 

\section{Results}

\label{sec:results}

\subsection{SEDs}

\begin{figure*}
\begin{center}
\includegraphics[width=0.8\textwidth]{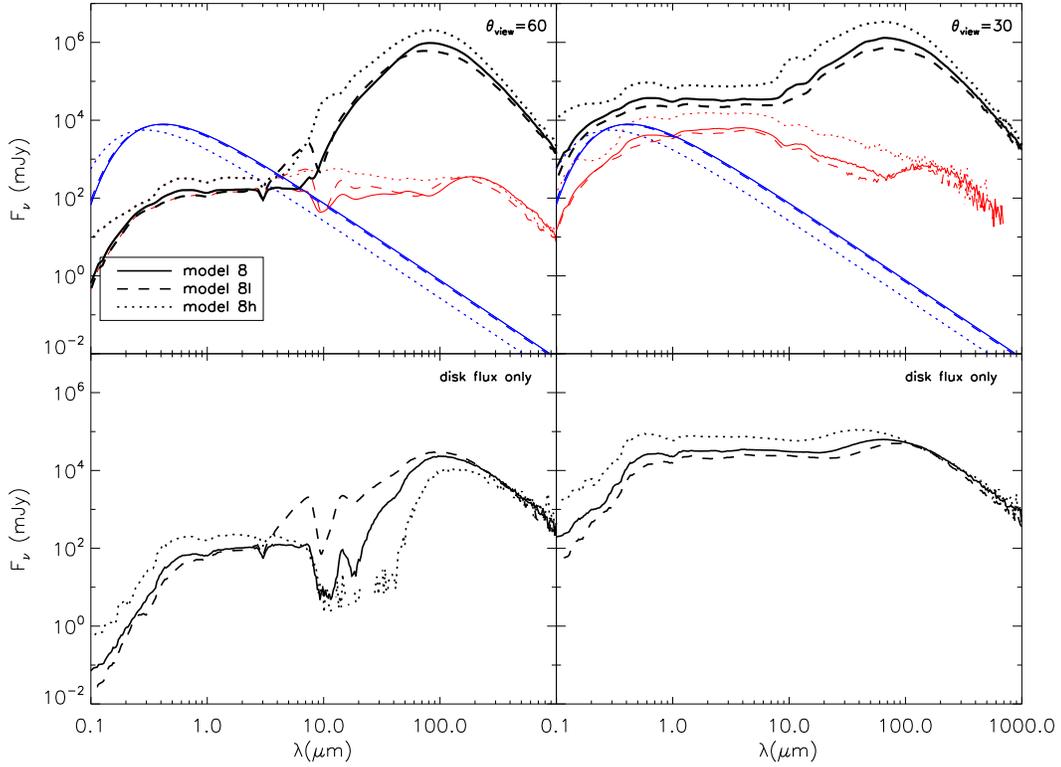}\\
\caption{SEDs of Model 8 ($\Sigma_\mathrm{cl}$= 1 g/cm$^2$), Model 8l
  ($\Sigma_\mathrm{cl}$= 0.316 g/cm$^2$) and Model 8h
  ($\Sigma_\mathrm{cl}$= 3.16 g/cm$^2$) at inclinations of $60^\circ$
  (left panels) and $30^\circ$ (right panels), assuming a distance of
  1 kpc. For each inclination, the upper panel shows the total fluxes
  (black), scattered fluxes (red) and stellar input (blue), while the
  lower panel shows fluxes from the disk only.}
\label{fig:sed_sigma}
\end{center}
\end{figure*}

\subsubsection{SEDs of the Model Series}

Figures \ref{fig:sed_series_60} and \ref{fig:sed_series_30} show the
SEDs of the model series at inclination angle between our line of
sight and the protostar rotation axis of $60^\circ$ (i.e. a more
equatorial view) and $30^\circ$ (i.e. a more polar view)
respectively. In each figure, the upper panel shows the total fluxes
while in the lower panel these SEDs are compared to the fiducial model
(Model 8). The $\theta=60^\circ$ line of sight goes through the
envelope material, in which case the photons from the protostar and
the disk are all reprocessed by the dust before they can escape the
core. The $\theta=30^\circ$ line of sight from the central star passes
through the outflow cavity (for Model 4 - 8), in which case we can see
the stellar black-body spectra in the short wavelength region of the
SEDs. Some important features in the SEDs include the water ice
feature at 3 {\micron} and silicate features at 10 {\micron} and 18
{\micron}. The ice feature is only present for the higher inclination
view for which the lines of sight pass through the envelope material,
which uses a dust model with ice mantles.

Models 1 to 3 show very similar SEDs, where we do not see any
radiation at short wavelengths. The 10 {\micron} silicate features are
all very deep. The occurrence of the expansion wave (Model 2) and
rotation/disk (Model 3) shift the far-IR peak a little to shorter
wavelengths, and increases the mid-IR emission, making the 18
{\micron} silicate feature deeper. This is because the expansion wave
and the Ulrich solution decrease the density of the inner region of
the core, and thus reduce the extinction. For Model 3, which is not
spherically symmetric, the SEDs do not show much difference between
the different inclinations. It should be noted that in Model 3
rotation is only considered inside the sonic point. In a more
realistic solution, the material in the outer region of the envelope
would also be redistributed by the effect of the rotation (like the
solution of \citet[]{TSC84} for an isothermal core) so that one might
see larger differences.

The outflow cavities change the shape of SEDs significantly. With
outflow cavities (Model 4 to 8), the SEDs show a large excess at
wavelengths shorter than 10 {\micron}. The position and height of the
far-IR peak and the 20 {\micron} $\sim$ 70 {\micron} slopes are
affected by the cavity as well. Especially for a low inclination, the
star and disk can be seen directly. The fluxes at wavelengths shorter
than 10 {\micron} are larger than those at a high inclination by about
two orders of magnitude.
Note that the short wavelength emission seen in the $\theta=60^\circ$
view is essentially all due to scattered emission from the outflow
cavity walls.

Compared to a passive disk, an active disk with accretion luminosity
increases the fluxes at all wavelengths without many changes in the
shape of the SEDs (comparing Model 4 and 5). The accretion luminosity
in Model 5 is mainly due to the hot-spot, while most of the disk
accretion luminosity is lost here because of the absence of the
innermost disk. The total energy is conserved only in Model 8, which
the disk is extended to the stellar surface.

The effect of the thickness of the disk is distinct on the SEDs
(comparing Model 5 and 6). A thicker disk tends to obscure more
photons at high inclinations and emit or scatter them to low
inclination directions - i.e. more flux escapes via the outflow
cavities. Therefore, with a thicker disk,
Model 6 at $\theta=30^\circ$ shows a rise between 1 and 10 {\micron}
which even smooths out the far-IR peak, while at $\theta=60^\circ$ the
SED shows a decrease in near-IR and shorter wavelengths, and a deeper
silicate features.

Model 8 shows how the SED changes by including the flux from the
innermost disk region. Compared to previous models, the optical and
near-IR emission is significantly increased, in both high and low
inclinations. Model 7 has exactly the same geometry and density setup
as Model 8, except that it has a disk truncated at
$r_\mathrm{sub}$. The opacities are chosen depending on the input
temperature and density in Model 8 while in Model 7 they only depend
on the density, therefore, even outside $r_\mathrm{sub}$ the opacity
setup of these two models can be different. Because of the truncated
disk, the majority of the disk accretion luminosity is lost in Model
7. This missing luminosity shows up in Model 8 mainly as optical
radiation. However, the near-IR flux in Model 8 is not so bright as
Model 7. The far-IR SEDs of these two models do not show much
difference.

Fig. \ref{fig:tdiffus_78} shows the final output temperature
distributions of these two models. The temperature of the disk reaches
$\sim10^5$ K at the innermost region in Model 8, while the highest
temperature in Model 7 is only $\sim 3000$ K. This explains the optical
excess in the SED of Model 8. However, the existence of the innermost
disk shields the flux from the star, therefore in Model 7 the
temperature at the surface of the disk and the base of the outflow
cavity becomes higher, leading to the higher near-IR flux on the
SED. For the models presented here, we do not have any material in the outflow
cavity. If any dust grain exists there, the optical emission in Model
8 will be suppressed and the disk luminosity may appear as more near-
and mid-IR radiation. This will be examined in a future paper.

\subsubsection{SEDs of the Fiducial Model}

Fig. \ref{fig:sed_fiducial} shows the SEDs of our fiducial model
(Model 8) at four inclinations. Even after we smoothed the
monochromatic opacity curves of gas, the original output SEDs still
show some significant emission features, mainly H$\alpha$. Since the
line profile is not exactly correct in our models and it is not the
interest of our present study, we subtract the H$\alpha$ feature
and smooth out other line features with wavelengths $<2$
{\micron}. The energy contained in the H$\alpha$ line is 
typically $\lesssim$ few\%.

The SEDs at 4 inclinations are shown here, from a view along the
rotation/outflow axis to that through the disk plane. The inclination
of the viewing angle changes the observed flux at short wavelengths
significantly. It also affects the height and position of the peak in
the far-IR region, and the mid-IR spectral slope. The SED at
wavelengths longer than $\sim$100 {\micron} is not affected by the
inclination. Inclinations of $30^\circ$ and $0^\circ$ have very
similar SEDs ($0^\circ$ inclination contains 8\% more energy than
$30^\circ$ inclination). Recall, the outflow cavity has an opening
angle of $51^\circ$, which means we can directly see the star in both
these cases. The stellar spectrum and the black-body spectrum
containing both stellar and hot-spot luminosity are shown by the
dotted lines. Their difference shows the luminosity from the hot-spot
region. From high inclinations to low inclinations, because of the
change of the optical depth, the silicate feature change from a big
absorption feature to a weak emission feature.

The dashed lines show the SEDs of only scattered light. At high
inclination angles, the observed light at short-wavelengths has always
been scattered, while at low inclinations, we can see stellar
radiation and thermal emission from the disk. In the far-IR, the flux is
dominated by the thermal emission of the envelope. Such significant
difference between low inclinations and high inclinations is partly
because we have an empty outflow cavity. 

\subsubsection{Effect of Different Mass Surface Density}

\begin{figure*}
\begin{center}
\includegraphics[width=0.8\textwidth]{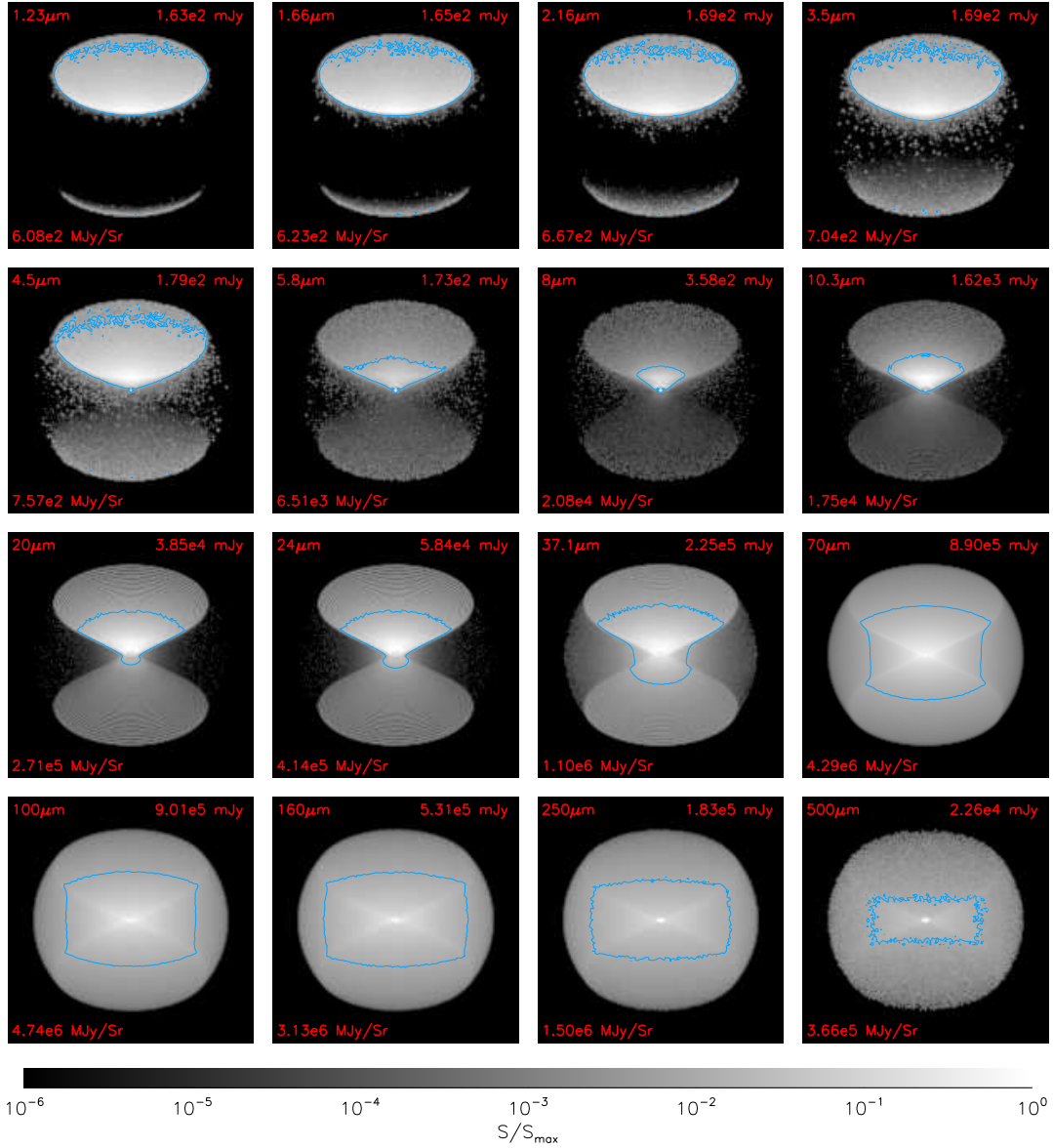}\\
\caption{Images for the fiducial model at inclination of $60^\circ$,
  assuming a distance of 1 kpc, at different wavelengths. Each image
  has 149 $\times$ 149 pixels, and field of view of 30" $\times$ 30".
  Each image is normalized to its maximum surface brightness which is
  labeled in the bottom-left corner. The values on the top-right
  corners are the total fluxes. The blue contour is 1\% of the maximum
  surface brightness}
\label{fig:img_60}
\end{center}
\end{figure*}

\begin{figure*}
\begin{center}
\includegraphics[width=0.8\textwidth]{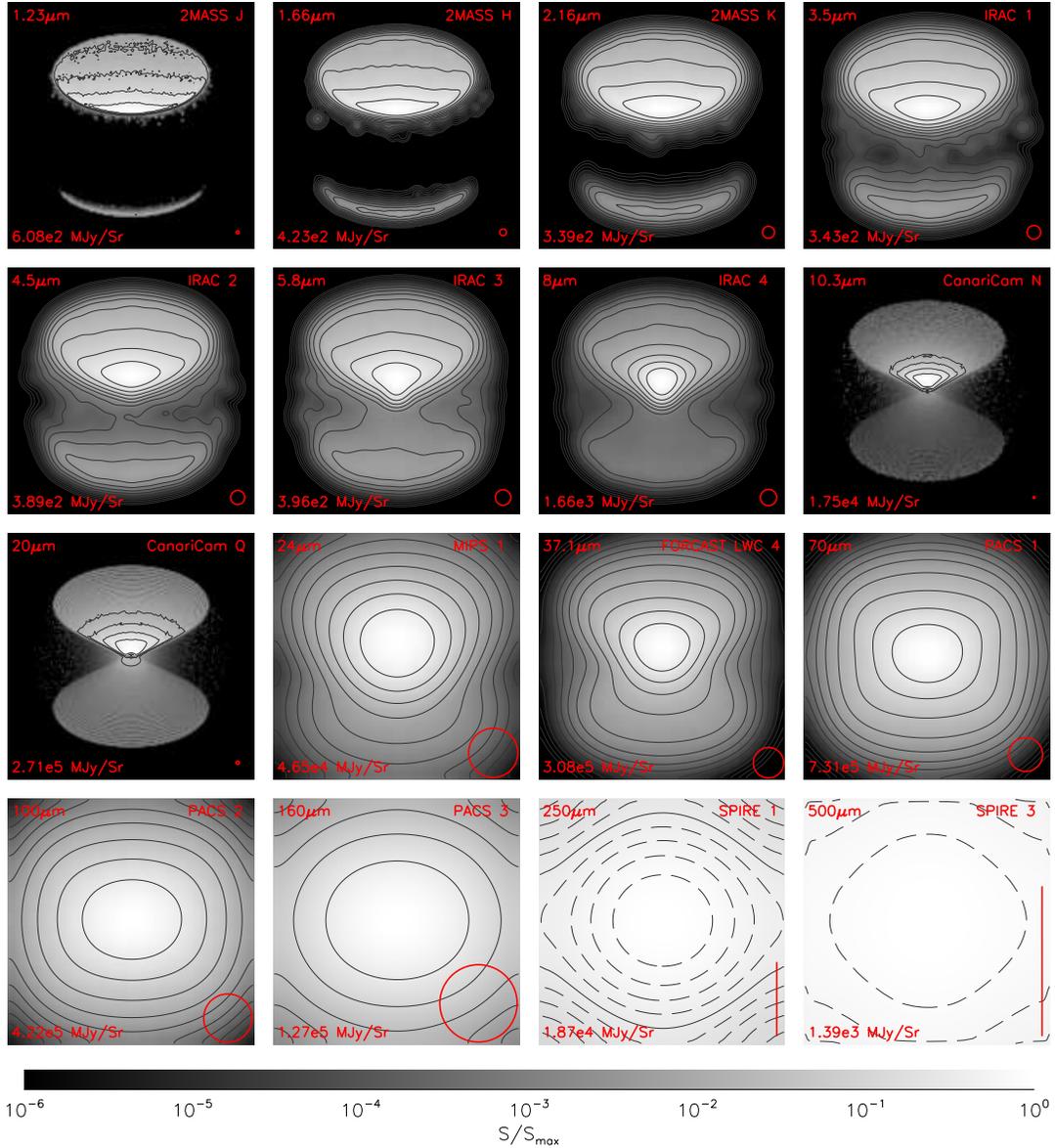}\\
\caption{Same as Fig. \ref{fig:img_60}, except convolved with the
  resolution of each particular instrument, labeled in the top-right
  corner and shown as red circles of 
radius equal to the HWHM (or line segments with lengths equal to the
HWHMs for Herschel SPIRE) on the bottom right corners. The maximum
surface brightness (which is dependent on the resolution) is labeled
in the bottom-left corner. The contours for most of the images have
intervals of half an order of magnitude from the maximum
brightness. Because some images (1.23, 10.3 and 20 {\micron})
are relatively noisy at faint fluxes, only five contours with
intervals of half an order of magnitude are shown here. For images at
250 and 500 {\micron}, dashed contours have intervals of 0.1 dex.}
\label{fig:img_60_conv}
\end{center}
\end{figure*}

\begin{figure*}
\begin{center}
\includegraphics[width=0.8\textwidth]{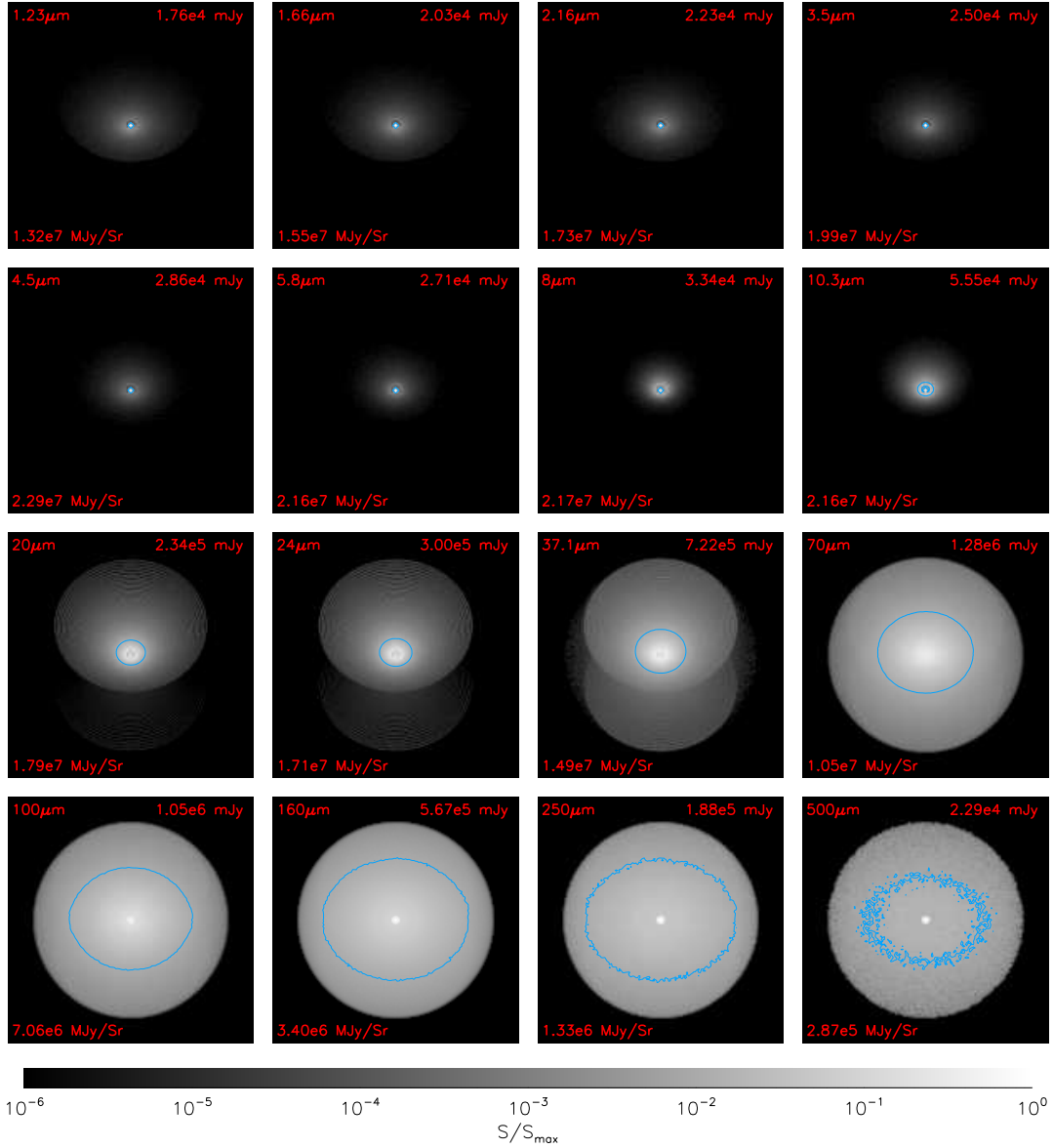}\\
\caption{Same as Fig. \ref{fig:img_60}, except at inclination of $30^\circ$.}
\label{fig:img_30}
\end{center}
\end{figure*}

\begin{figure*}
\begin{center}
\includegraphics[width=0.8\textwidth]{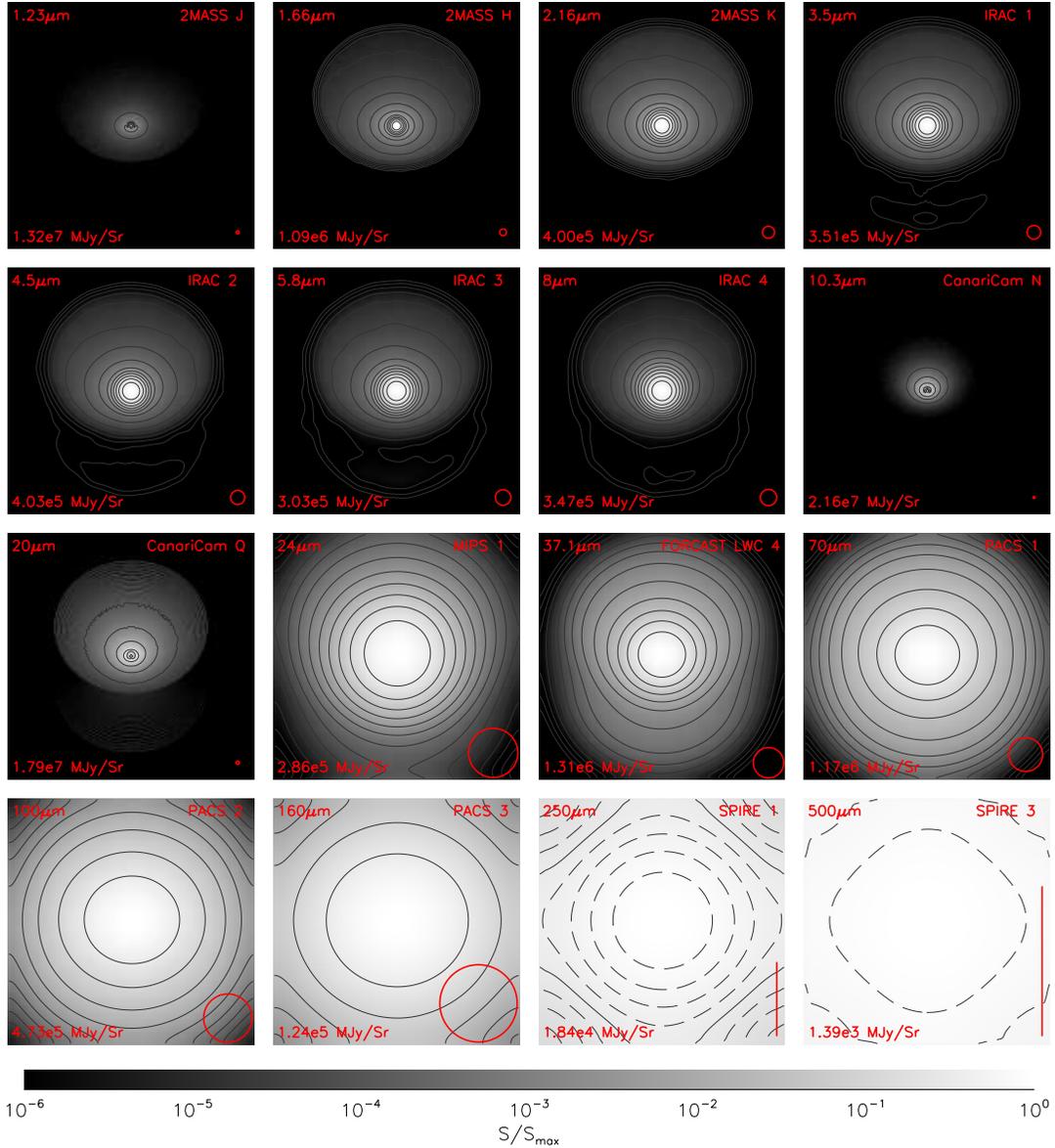}\\
\caption{Same as Fig. \ref{fig:img_60_conv}, except at inclination of $30^\circ$ and that for images at 1.23, 10.3 and 20 {\micron}, only five contours with intervals of an order of magnitude are shown.}
\label{fig:img_30_conv}
\end{center}
\end{figure*}

\begin{figure}
\begin{center}
\includegraphics[width=\columnwidth]{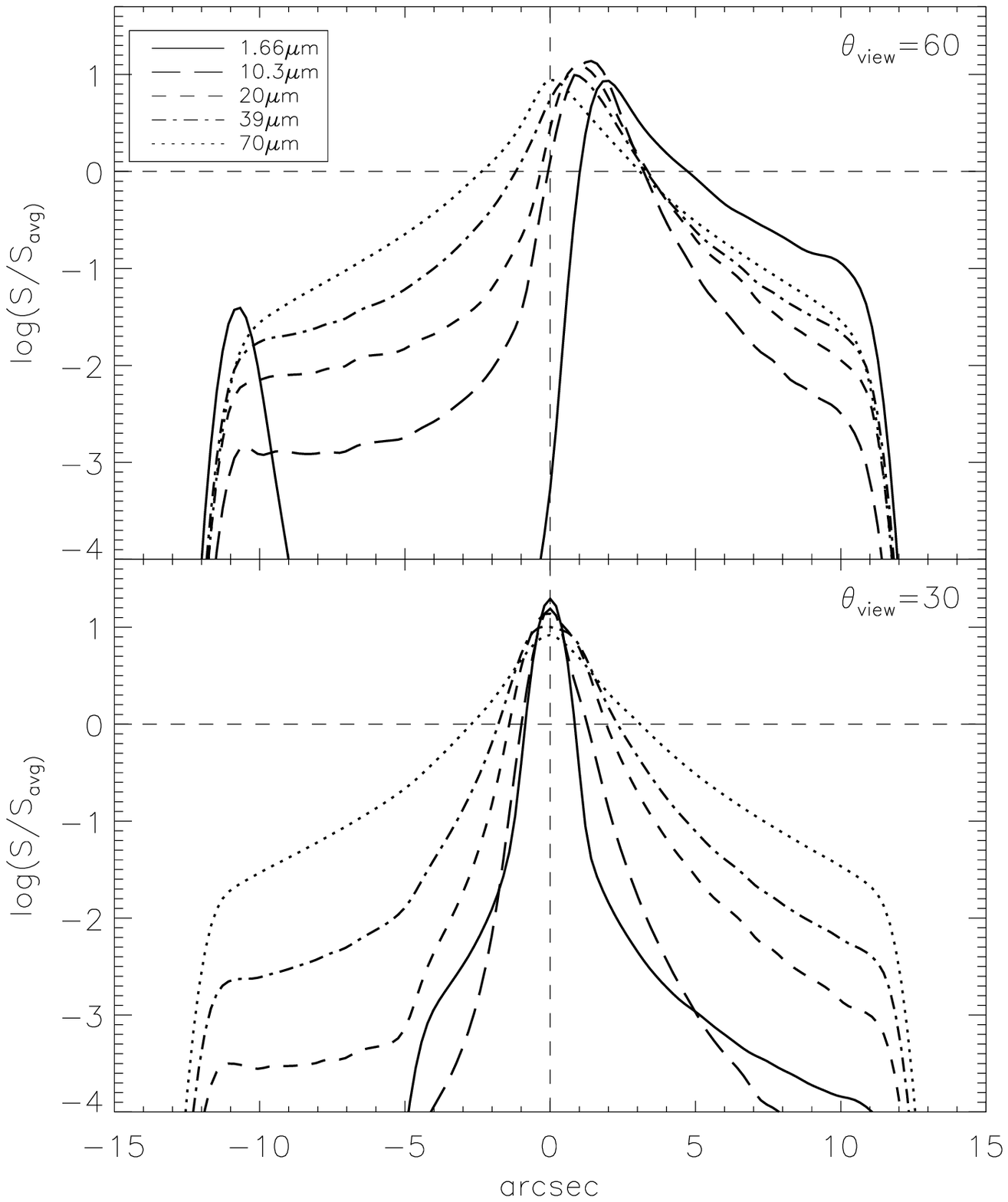}\\
\caption{Flux distribution along the projected outflow axis. Fluxes
  are normalized by the mean value of a narrow strip across the core
  along the axis. Different curves show different wavelengths. The
  upper panel is at inclination of $60^\circ$ and the lower panel is
  at $30^\circ$.}
\label{fig:fluxdist}
\end{center}
\end{figure}

Model 8, 8l and 8h compare the effect of different mean surface
densities of clump in which the core is embedded, which affects the
surface pressure of the core, and therefore the size of the core, the
accretion rate, the disk structure and the protostar evolution. The
size of the core, the radius of the expansion wave front, the position
of the sonic point and the disk radius are all proportional to
$\Sigma_\mathrm{cl}^{-1/2}$, so with higher clump surface density, the
core is more compact and accordingly the accretion rate is higher. The
scale hight of the disk is also larger in the high
$\Sigma_\mathrm{cl}$ case. The stellar luminosities of these three
models are similar but Model 8h has a much bluer stellar spectrum,
because of the higher hot-spot accretion luminosity.
Some important parameters of these three models are listed in
Table. \ref{table2}

Fig. \ref{fig:sed_sigma} compares SEDs of these three models at both
$60^\circ$ and $30^\circ$. As discussed above, higher surface density
leads to higher bolometric luminosities, which can be seen from the
SEDs: At inclination of $30^\circ$, with a higher surface density, the
flux is higher at all wavelengths; And at $60^\circ$ inclination we
can see the same effect in optical, near-IR and far-IR
emissions. However, in mid-IR, Model 8l shows a rise of the flux and a
very significant 10 {\micron} silicate feature, while the other two
models do not. To explain this, we also show the disk flux in the
lower panels. Here, disk flux contains photons which have their last
emission in the disk, and then either escape the core directly or are
scattered before they reach the observer.  At the inclination of
$60^\circ$, the disk is blocked by the envelope. The short-wavelength
fluxes from the disk should all have been scattered. They keep the
trend that the model with higher surface density has higher
fluxes. However, in mid- and far-IR, the fluxes should have suffered
the extinction of the envelope, making the model with higher surface
density have lower fluxes because of the higher extinction. Thus, even
though Model 8 and 8h have very strong silicate features in their disk
SEDs, they are buried in the envelope fluxes in the total SEDs, while
Model 8l shows the high disk flux level in mid-IR and a significant
silicate absorption feature because of its lower extinction and lower
envelope flux.

\subsection{Images}

\subsubsection{Images of the Fiducial Model}

Fig. \ref{fig:img_60}, \ref{fig:img_60_conv}, \ref{fig:img_30} and
\ref{fig:img_30_conv} show the images for the fiducial model (Model 8)
at inclinations of $60^\circ$ and $30^\circ$, assuming a distance of 1
kpc and no foreground extinction. The instruments filters chosen here
are: 2MASS J, H, K bands, Spitzer IRAC 3.5 {\micron}, 4.5 {\micron},
5.8 {\micron} and 8 {\micron}, MIPS 24 {\micron}, GTC-CanariCam N band
and Q wide band, SOFIA FORCAST 37.1 {\micron}, Herschel PACS 70
{\micron}, 100 {\micron}, 160 {\micron}, and SPIRE 250 {\micron} and
500 {\micron}.  We show both resolved images and those after
convolving with PSFs of the particular instruments. Images showed here
are all normalized to their maximum surface brightnesses which are
labeled on each image. $10^9$ photon packets are used. However they
are still not enough to produce a smooth image for the narrow filters
or at the wavelengths with low fluxes. Besides, the simulation grid
also contributes to this problem, like the strip patterns in the
images at some wavelengths (e.g. 20 {\micron}).  Around the opening
angle of the outflow cavity, the grid with very small change of the
polar angle intersects with the cavity wall at quite different
radii. Even we have made the grid much finer in polar angle at the
region around the cavity wall, these patterns can still be seen,
especially at outer regions of the envelope. A finer grid would demand
more memory and computing time.

On the images at inclination of $60^\circ$ before convolution with the
instrument PSF, the most significant features are the outflow
cavities. They can be seen in any wavelength, though they become not
so obvious in far-IR wavelengths where the thermal emission of the
envelope dominates. At this inclination, the line of sight intersects
with the envelope, thus in near-IR the only emission we can see is
scattered by the cavity wall and escapes from the opening region,
especially from the side facing us. Deeper regions appear as the
wavelength increases. The central protostar and disk begins to show up
in mid-IR images. The dark lane around it shows the size of the
disk. In the mid-IR, the cavity walls dominates the emissions, as has
been discussed by \citet[]{deBuizer06}. Both sides can be clearly seen
and the brightest region is the base of the cavity. The emission of
the envelope takes over at longer wavelengths, making the image
symmetric and the outflow cavity begins to fade. The central star and
disk can still be seen as the brightest region in those wavelengths.

At the inclination of $30^\circ$, the central object can always be
seen through the empty outflow cavity. At shorter wavelengths, only
the side facing us of the cavity is significant and the opposite side
is very dim. Far-IR emission is dominated by the envelope and it is
hard to tell the features of the cavities. It should be noted here,
especially when comparing the 30$^\circ$ and 60$^\circ$ images, that
the images are all scaled to their maximum surface brightness, which
is generally much greater for the 30$^\circ$ viewing angles.

After convolving with PSF of particular instruments, in the far-IR,
the contours become very symmetric and we cannot tell the inclination
or the opposite outflow cavities. It is possible to see the opposite
outflow in MIPS 24 {\micron} and FORCAST 37.1 {\micron} at high
inclinations, if the S/N is large enough.  The two sides of the
outflow cavities are clear in IRAC images and in near-IR
images. GTC-CanariCam has very high resolution so it may enable us to
see the central region in mid-IR.

Fig. \ref{fig:fluxdist} shows the flux profile along the system axis
at inclinations of both $60^\circ$ and $30^\circ$. Each profile is
normalized by the mean flux of a very narrow strip along the axis
across the whole core. Thus, the general shape and level of the
profiles are independent on the resolution of the image. But the curve
would be smoothed out if a larger PSF-FWHM is used. At inclination of
$60^\circ$, as we discussed above, at short wavelengths, the central
region is totally blocked by the envelope, most of the emission comes
from the outflow cavity. In the mid-IR, the maximum surface brightness
comes from the base of the outflow cavity rather than central star and
disk, on the side the outflow cavity is opposite to us, the flux drops
very fast by almost four orders of magnitude, while on the side the
outflow is facing us the flux drops much more gradually. In the
far-IR, the profile is symmetric on both sides, and central region has
the maximum surface brightness. At lower inclination, the maximum flux
always comes from the center. In the far-IR, the profile is very
similar to the case in higher inclination. At shorter wavelengths, the
contrast between outer regions and the center point is much higher
than the previous case, so that the profile drops very fast. At 1.66
{\micron} and 10.3 {\micron} we can only see the flux due to the
outflow cavity facing us, while the other side only shows up at 20
{\micron}. Because in our models it is empty outside the core, the
profiles all cut off at the core radius, which is not true in reality
since clump material can extend to larger radii.

\subsubsection{Effect of Different Surface Density}

\begin{figure*}
\begin{center}
\includegraphics[width=0.8\textwidth]{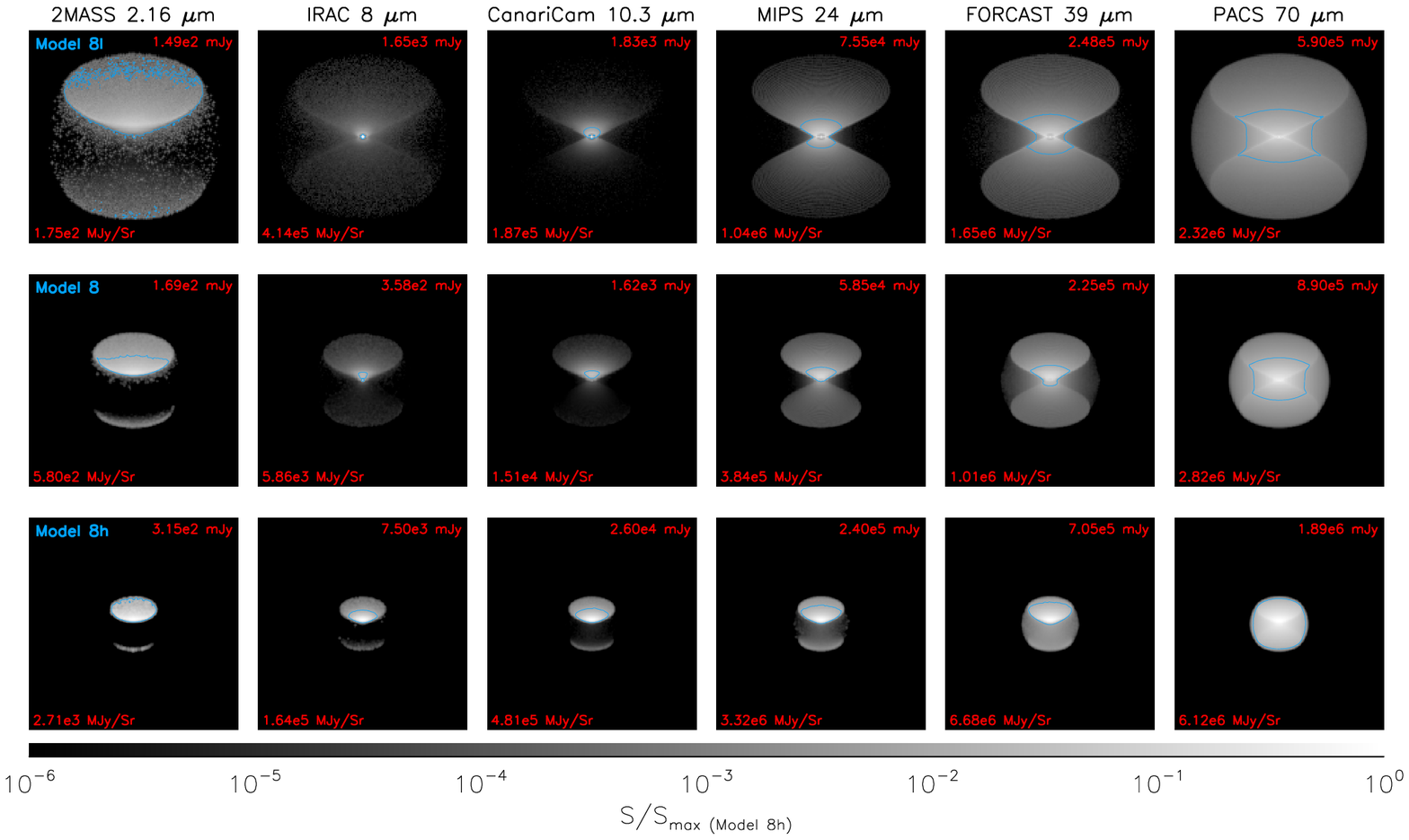}\\
\caption{Images for Model 8, Model 8l and Model 8h at the inclination
  of $60^\circ$, assuming a distance of 1 kpc. The size of the image
  is 50" $\times$ 50". The resolution here is set to be 0.5", so the
  surface brightnesses can be compared. Each images are scaled to the
  maximum brightness of Model 8h at that band, but their own maximum
  brightness is also labelled in the bottom-left corners. The total
  flux in each image is also labelled in the top-right corner. The
  blue contour is 1\% of the maximum surface brightness in that
  image.}
\label{fig:img_sigma_60}
\end{center}
\end{figure*}

Fig. \ref{fig:img_sigma_60} compares the images of Model 8, 8l and 8h
at 6 wavelengths. The size of the images are all 50"$\times$50". We
can see that the core is smaller when the surface pressure is
higher. The surface brightness is dependent on the resolution, so here
we convolved all images with same PSF with FWHM of 0.5". The images
are scaled to the maximum surface brightness of Model 8h at each
wavelengths, so the brightnesses of the three images at each
wavelength can be directly compared.

The total fluxes and the maximum surface brightness in images of Model
8h are generally higher than those of the other two models. This is
natural since the total luminosity is much higher in this model. The
optical depth is larger when $\Sigma_\mathrm{cl}$ is
higher. 
In the near-IR, we can see the base of the outflow
cavity facing us and most of the opposite side in Model 8l, while we
can only see the light from the opening region of the cavities in
Model 8h. The central region shows up in Model 8h only in
far-IR, while it can be seen at mid-IR in Model 8 and 8l. With lower
surface density, the contrast between two sides of the outflow
cavities becomes smaller. In Model 8h the side towards us is much
brighter than the other side, even at 70 {\micron}, while the other
two models show almost symmetric images. 
We can also see that the mid-IR emission is dominated by the cavity walls,
especially in the cases with higher extinction, as opposed to that with lower extinction
the emissions are more concentrated to the central region.

\section{Discussions and Conclusions}

\label{sec:conclusions}

We have constructed a model for individual massive star formation,
with a $60\;M_\odot$ initial core that forms an $8M_\odot$ protostar. We included the
inside-out expansion wave in the core, free-fall rotating collapse in
the inner region, an accretion disk of 1/3 the mass of the star, and
bipolar outflow cavities with large opening angle, parameters of which
are all calculated self-consistently. For the first time, we
considered an optically thick inner disk with gas opacities which were
assigned depending on the pre-calculated temperature profile. This
inner disk enabled us to calculate the emission from the active disk
with correct total luminosity and spectrum.

Compared to the Robitaille model, our parameter space covers higher
accretion rates, higher disk mass, denser envelopes and larger outflow
cavities.  Also, in the Robitaille model the disk accretion luminosity
is much lower than that in our model since the disk is truncated at
the dust destruction radius, while the hot-spot luminosity on the
stellar surface is set to be $Gm_*\dot{m}(1/r_*-1/r_\mathrm{co})$
where $r_\mathrm{co}$ is the magnetic co-rotation radius and set to be
$5r_*$, making the hot-spot luminosity is 4/5 of the total accretion
luminosity $L_\mathrm{acc}$. So energy of
$\lessapprox0.2\;L_\mathrm{acc}$ is lost. In our model, the total
accretion luminosity is divided equally into disk accretion luminosity
(emitted from the disk extended to stellar radius with gas and dust
opacities) and hot-spot luminosity (added on the stellar
spectrum). Also, the Ulrich solution is used for the whole envelope in
Robitaille model, which means the whole envelope is assumed to be
free-falling and the infalling rate is constant at all radii. As
discussed in Section \ref{sec:rotating}, the core undergoes free-fall
only in the inner region. Therefore, for a given core mass and
accretion rate, our model has a more compact core with higher
extinction, causing the far-IR peak of the SED to be at lower fluxes
and at longer wavelengths.

We have presented SEDs of the model series, the fiducial model, and
the models with higher and lower surface pressure, at typical
inclinations. We also have presented images for the fiducial model at
JHK bands, IRAC and MIPS bands, GTC-CanariCam bands, SOFIA bands and
Herschel bands, both resolved and convolved with the resolutions of
each instrument. The main conclusions can be summarized as:

1. Outflow cavities affect the SEDs significantly and cause a large
difference between low and high inclinations of our viewing
angle. However, the present modeling assumes these cavities are
optically thin, which may not be valid, especially at the shorter
wavelengths. This issue will be investigated in a follow-up paper.
The height of the disk also affects the SEDs. With a thicker
disk, the near- and mid-IR fluxes at low inclinations become higher,
while at high inclinations it suppresses the fluxes (especially at
short wavelengths) and creates deep silicate features. Also, the
density distribution in the core (especially the inner region) can
affect the mid-IR flux levels, the silicate features, and the
far-IR peaks.

2. The temperature of the innermost region of the disk can reach $\sim
10^5$ K. The disk becomes optically thick in such conditions even if
no dust can exist there. SEDs show the rise of optical emission due to
this hot inner disk. This optically thick inner disk also shields flux
from the protostar, leading to lower temperatures on the surface of
the disk and the base of the cavity wall, and therefore lower fluxes in
near-IR and mid-IR part of the SEDs.

3. The SEDs of the fiducial model show that the inclination can affect
SEDs at wavelengths shorter than 100 {\micron}, including the far-IR
peak. The mid-IR spectral slope changes significantly with
inclination. The silicate feature changes from a deep absorption
feature to an emission feature from high inclinations to low
inclinations, due to the change of the optical depth.

4. With higher surface pressure, the core becomes more compact and the
accretion rate and luminosity become higher, leading to higher fluxes
at all wavelengths (except in the mid-IR for high inclination cases
which suffer higher extinction).
High extinction can also cause the mid-IR flux to be dominated by the
envelope, and thus hide the silicate absorption feature. Thus, only in
the model with low extinction does the silicate absorption feature
show up.

5. Outflow cavities are the most significant features on the images,
except at wavelengths longer than 70 {\micron}. At inclinations of
$60^\circ$, from short wavelengths to long wavelengths, the brightest
point moves from the outer region of the cavity to the base of the
cavity wall, and to the center of the core in far-IR, while at
inclination of $30^\circ$, it is always the central region. At
inclination of $60^\circ$, the opposite outflow cavity can be seen 
n the mid-IR if fluxes $\sim 10^2 - 10^3$ times fainter than the peak
can be probed. It is very difficult to see the opposite cavity at an
inclination of $30^\circ$. GTC-CanariCam (and other 10 m class
telescopes with mid-IR cameras) has very high angular resolution so it
may enable us to resolve the central disk system. The flux distribution
along the outflow axis can help constrain model assumptions and
inclination angles and so will useful to measure.
In the mid-IR, the cavity walls seem to dominate the emission, but for
lower density cores with lower extinction, the central region becomes
brighter.  The contrast between the two sides of the outflows in mid-
and far-IR increases as the extinction becomes higher.

The model we have presented will be improved in future papers by a
detailed consideration of the effect of including gas and dust in the
outflow cavity. Currently models for the density distribution here are
quite uncertain, which is why we defer this study to a future date. An
evolutionary sequence of protostellar models with and without outflow
opacity will then be presented. To gauge the degree of inhomogeneity
we will consider the results of numerical simulations of core
accretion and outflow interaction (e.g. \citealt[]{Krumholz07},
\citealt[]{Staff10}), and then incorporate these inhomogeneities into
the models in a parametrized form.

\acknowledgements 

We thank James De Buizer and Barbara Whitney for helpful discussions.
YZ acknowledges support from an Alumni Fellowship of the University of
Florida.  JCT acknowledges support from NSF CAREER grant AST-0645412;
NASA Astrophysics Theory and Fundamental Physics grant ATP09-0094; and
NASA Astrophysics Data Analysis Program ADAP10-0110.

\bibliography{zhang_2011_apj}

\end{document}